\documentclass[]{JFM-FLM_Au}
\usepackage{subcaption}
\usepackage{multirow}
\usepackage{xcolor}

\lefttitle{C. Lin and S. Liao}
\righttitle{Journal of Fluid Mechanics}

\title{Convergent series of Stokes wave of arbitrary height in deep water via machine learning}

\author{Chong Lin\aff{2} \and Shijun Liao\aff{1,2}}

\affiliation{\aff{1}State Key Laboratory of Ocean Engineering, Shanghai 200240, China
\aff{2}School of Ocean and Civil Engineering, Shanghai Jiao Tong University, Shanghai 200240, China}

\corresau{Shijun Liao, \email{sjliao@sjtu.edu.cn}}

\begin{document}
\maketitle

\begin{abstract}
Permanent gravity waves propagating in deep water, spanning amplitudes from infinitesimal to their theoretical limiting values, remain a classical yet challenging problem due to its inherent nonlinear complexities. Traditional analytical and numerical methods encounter substantial difficulties near the limiting wave condition due to singularities at sharp wave crests. In this study, we propose a novel hybrid framework combining the homotopy analysis method (HAM) with machine learning (ML) to efficiently compute convergent series solutions of Stokes waves in deep water for arbitrary wave amplitudes from small to theoretical limiting values, which show excellent agreement with established benchmarks.  We introduce a neural network trained using only 20 representative cases whose series solution are given by means of HAM, which can rapidly predict series solutions across arbitrary steepness levels, substantially improving computational efficiency. Additionally, we develop a neural network to gain the inverse mapping from   the conformal coordinates $(\theta, r)$ to the physical coordinates $(x,y)$, facilitating explicit and intuitive representations of series solutions in physical plane.   This HAM-ML hybrid framework represents a powerful and efficient approach to compute convergent series in a whole range of physical parameters for  water waves with arbitrary wave height including even limiting waves.    In this way we establish a new {\bf paradigm} to {\bf quickly} obtain convergent series solutions  of complex nonlinear systems for a whole range of physical parameters,  thereby significantly broadening the scope of series solutions that can be easily gained by means of HAM even for highly nonlinear problems in science and engineering.
\end{abstract}

\begin{keywords}  Stokes waves, arbitrary wave height, convergent series, machine learning

\end{keywords}


\section{Introduction}
\label{sec:intro}

The investigation of two-dimensional, steady, traveling gravity waves on the surface of an incompressible, inviscid fluid of infinite depth represents one of the most fundamental and enduring problems in fluid mechanics. These periodic waveforms, often referred to as Stokes waves, serve not only as canonical solutions in the theoretical study of nonlinear wave dynamics but also as essential building blocks in modelling more complex oceanic phenomena, including wave-wave interactions, wave breaking, and wave-flow coupling. Despite their seemingly simple appearance, the underlying governing equations are fully nonlinear and analytically intractable in general, particularly for waves of large amplitude. As such, this topic has remained a subject of intensive theoretical and computational investigation for over two centuries.

Classical approaches to analyzing periodic gravity waves date back to \citet{stokes1847oscillatory, stokes1880supplement}, who developed a perturbation expansion in terms of a small parameter \(a_1\), the first Fourier coefficient. Through this framework, he conjectured the existence of an ultimate limiting wave form, characterized by a sharp crest forming an internal angle of $120^\circ$. \citet{stokes1880supplement} computed the solution to \(\textit{O}(a_1^5)\) in the special case of infinite depth. \citet{wilton1914deep} carried the infinite depth computation up to \(\textit{O}(a_1^{10})\) but has errors starting with his eighth-order results. Significant progress was made by \citet{schwartzComputerExtensionAnalytic1974a}, who noted that the first Fourier coefficient \(a_1\) does not increase monotonically with wave height. Instead, he introduced half wave height as a more suitable perturbation parameter, which enabled him to numerically compute high-order solutions up to the 117th order in deep water. His results revealed that all Fourier coefficients exhibit similar non-monotonic trends and confirmed the divergence of the classical Stokes series beyond a critical steepness. To address this issue, he employed Pad\'e approximants to extend the convergence range of the series in the vicinity of the limiting wave. \citet{schwartzComputerExtensionAnalytic1974a} found that accurate wave profiles cannot be obtained for very high waves even with the aid of the Pad\'e approximant. 

Subsequent developments sought to improve the convergence of perturbation series by redefining the perturbation quantity. \citet{longuet-higgins1974solitary} and \citet{longuet-higgins1975integral} respectively restated the perturbation series with more suitable perturbation quantities. This adjustment significantly improves the convergence characteristics of wave steepness in a wider range. More recently, \citet{dallaston2010accurate} revisited these classical approaches using high precision numerical techniques to minimize rounding errors. \citet{zhaoStokesWaveSolutions2022a} developed a new set of fifth-order Stokes wave solutions, using MATLAB Symbolic Toolbox. Despite these advancements, the perturbation method remain limited by their inherent dependence on small parameters and the complexity of extending series beyond moderate nonlinearity. These limitations continue to motivate the development of alternative, non-perturbative analytical approaches.

To address the limitations of classical perturbation theory, researchers have developed a lot of advanced numerical methods over the past several decades. \citet{chappelear1961direct} use a least-squares approach by expanding the free surface and velocity potential in Fourier series, then determining the coefficients by the method of least squares. \citet{dean1965stream} extended this concept by adopting an analytic streamfunction representation with unknown coefficients, which he computed via a numerical perturbation scheme. Building on this, \citet{williams1981limiting} proposed an improved integral formulation that incorporated a specialized crest-correction term. By enhancing convergence near the wave crest, this innovation allowed high accuracy with significantly fewer Fourier modes. In parallel, \citet{rienecker1981fourier} introduced a Fourier-based method, reducing the nonlinear boundary conditions to a finite set of algebraic equations solved via Newton iteration. \citet{fenton1988numerical} streamlined the implementation by numerically evaluating the required derivatives, though convergence at large steepness often required stepwise extrapolation from smaller wave heights. The search for more extreme solutions led \citet{Lukomsky2002} to investigate sharp-crested deep-water waves exhibiting local singularities within the flow domain. \citet{maklakov2002almost} proposed a method to calculate periodic water waves based on solving integral equations, and the calculated wave steepness can reach \(99.99997\%\) of the theoretical limiting wave. Recent studies that combine spectral discretization with Petviashvili-type method have broadened the numerical toolkit for steady traveling waves, including gravity and capillary-gravity regimes. \citet{Clamond2013fast} introduced a fast and accurate scheme for solitary gravity waves computation, and \citet{Dutykh2014efficient} further emphasized spectral accuracy and computational efficiency for steady solitary solutions. \citet{Clamond2015plethora} presented a wide family of generalized traveling wave solutions that interpolate between solitary and periodic profiles using a spectral Petviashvili-type approach, while \citet{Dutykh2016efficient} designed an efficient algorithm for capillary gravity generalized solitary waves, including cases close to limiting configurations. Building on these advances, \citet{clamond2018accurate} constructed a fast pseudo-spectral Petviashvili-type solver in conformal coordinates that achieved high accuracy across arbitrary depths. Most recently, \citet{chen2024accuracy} revisited Fourier spectral approximation (FSA) methods and  evaluated the accuracy of the FSA for almost-limiting gravity waves.

In the study of deep-water waves, increasing attention has been directed toward the complex dynamics near the crest of limiting Stokes waves. \citet{longuet-higgins1978theory} proposed a matching technique to analyze gravity waves of nearly extreme form, thereby confirming the existence of branch points of order $1/2$, as previously predicted by \citet{grant1973singularity}, as well as the presence of turning points in the phase velocity as a function of wave height. These findings offer compelling evidence of the complex analytic structure near the limiting Stokes wave and underscore the fundamental difficulties associated with modeling such waves using standard perturbation methods. \citet{dallaston2010accurate} successfully predicted the dispersion relation and three turning points in the total energy. Building on this foundation, \citet{tanveer1991singularities} and \citet{crew2016new} used analytic continuation into the complex plane to uncover the nature of these singularities. \citet{lushnikov2016structure} further explored the global analytic structure of non-limiting and limiting Stokes waves using high-precision numerical simulations. He identified a square-root branch point on the physical (first) sheet of the Riemann surface for non-limiting waves, and an infinite number of such singularities residing on higher non-physical sheets. Based on these findings, \citet{lushnikov2016structure} conjectured that the $2/3$-power law singularity at the crest of limiting Stokes wave emerges from the coalescence of these nested square-root singularities. This hypothesis was strongly supported by numerical evidence. To improve convergence near the crest, \citet{lushnikov2017new} introduced an auxiliary conformal mapping with a tunable parameter in order to allow for finer resolution near the crest of the wave and faster convergence. 

The persistent non-analytic behavior near the crest of steep water waves continues to present a major obstacle for both analytical and numerical modeling of nonlinear wave phenomena. Traditional methods often rely on extrapolation methods like Pad\'e approximants to improve the convergence. To address these limitations, the homotopy analysis method (HAM), originally proposed by \citet{liao1992phd, liao2003beyond, liao2009notes, liao2010optimal, liao2011multiple, liao2012ham, Liao2022AAMM, Liao2026AAMM}, introduces a fundamentally different analytical framework. Unlike classical perturbation techniques, HAM does not rely on any small or large physical parameter. It introduces an auxiliary convergence control parameter, expressed by \( \hbar \), which can explicitly adjust the convergence of homotopy series solutions. This additional degree of freedom grants the method substantial flexibility, allowing it to construct rapidly converging series even in strongly nonlinear regimes where conventional approaches diverge. HAM has proven successful in a wide range of applications, including nonlinear wave problems, steady-state resonant wave problems, and other highly nonlinear mechanical systems~\citep{liao1999uniform, liao2003homotopy, vangorder2008laneemden, vajravelu2012nonlinear, liu2014resonance, liu2015existence, zhong2017analytic, fang2023homotopy, Li2024, Xie2025JCP}. More importantly, HAM has led to the discovery of an entirely new class of solutions, known as steady-state resonant waves. These highly nonlinear wave phenomena were first predicted theoretically using HAM by \citet{liao2011multiple, xu2012steady, liu2014resonance} and were subsequently confirmed through laboratory experiments~\citep{liu2015existence}. This not only highlights the capacity of HAM for handling strongly nonlinear problems but also underscores its potential for uncovering complex physical behaviors that remain inaccessible to other analytical approaches. Building on this foundation, \citet{zhong2018limiting} successfully applied HAM to construct convergent series solutions for limiting Stokes waves  of {\em arbitrary} depth including extremely shallow water. Their work demonstrated that even solitary waves in extremely shallow water, traditionally among the most challenging cases due to their strong nonlinearity and crest singularities, can be resolved with high precision using the HAM. 

Up to now, most approaches for nonlinear wave problems fail to establish a direct relationship between the wave steepness and the corresponding wave solution. Furthermore, they typically gain case-specific solutions based on discrete sets of physical parameters, lacking a unified formulation applicable to {\em arbitrary} wave conditions. As a result, any change in parameters requires repetitive and computationally expensive recalculations. More broadly, these limitations are not confined to water wave problems, but are common in many strongly nonlinear systems governed by nonlinear partial differential equations (PDEs).  Moreover, when physical parameters are retained symbolically, the resulting series solutions often become increasingly cumbersome, making it extremely difficult to ensure convergence in the presence of strong nonlinearity. Therefore, developing a general and efficient approach capable of rapidly constructing convergent series solutions for such nonlinear systems is of significant importance, as it greatly enhances the applicability and practicality of convergent series solutions. This is the motivation of this paper.

In recent years, machine learning has been increasingly applied to nonlinear physical problems, particularly in fluid dynamics, where it addresses key limitations of traditional methods and achieves state-of-the-art accuracy and efficiency, thereby making it possible to open a new pathway toward the analytical solution of complex systems~\citep{kolmogorov1991local, kutz2017deep, brunton2020machine, xie2020artificial, jouybari2021data, li2022machine, eeltink2022, lozano2023machine, xu2023artificial, lu2024physics, zhang2024filtered, zhu2024, wang2024,yousefi2024, chen2024, brechtPhysicsinformedNeuralNetworks2025}. Deep learning with deep neural networks (DNNs), a key subfield of machine learning, transforms input features through multiple layers of nonlinear operations~\citep{lecun2015deep}. A central advantage of DNNs lies in their architectural flexibility, which enables the incorporation of physical invariance principles such as tensor symmetry. In nonlinear wave problems, accurate representation of physical quantities often necessitates transformations between computational and physical coordinate systems. DNNs are well suited for learning these complex nonlinear mappings, ensuring physical consistency under operations like coordinate inversion, rotation, and reflection. These properties make DNNs particularly effective for modeling strongly nonlinear phenomena, where traditional numerical methods often suffer from high computational costs and limited generalizability. 

For example, \citet{ling2016reynolds} employed deep neural networks to model the Reynolds stress anisotropy tensor using high-fidelity simulation data. \citet{jouybari2021data} employed deep neural networks and Gaussian process regression to develop a high-fidelity predictive model for the Nikuradse equivalent sand-grain roughness height, $k_s$, in turbulent flows over a diverse range of rough surfaces. \citet{eeltink2022} integrated a physics-based nonlinear evolution model with a long short-term memory network to predict deep-water wave breaking and its impact. \citet{wang2024} presented a deep learning wave model that directly predicts global significant wave heights at high spatiotemporal resolution, generating one year of hourly predictions in under 30 minutes on a personal computer. \citet{chen2024} introduced WaveNets, a dual-network neural network that reconstructs velocity, pressure, and vorticity fields by embedding the full Euler equations into its loss function. Focusing on air-sea interaction, \citet{yousefi2024} proposed a supervised machine learning model to estimate the spatial distribution of skin-friction drag over wind waves. The model accurately predicts the overall distribution of viscous stresses, demonstrating strong agreement with high-resolution laboratory measurements. Meanwhile, \citet{brechtPhysicsinformedNeuralNetworks2025} employed neural networks to solve the shallow-water equations for tsunami modeling. Through several classical benchmark cases, they demonstrated that these networks can accurately capture shallow-water wave propagation and inundation dynamics, providing mesh-free and physically consistent solutions.

In this study, we propose a novel and general framework for analytically solving strongly nonlinear problems by integrating the homotopy analysis method with machine learning. Specifically, we first gain series solutions at discrete wave heights using homotopy analysis method, and then train a deep neural network to generalize these solutions for {\em arbitrary} wave heights. This enables, for the first time, {\em convergent} series solutions for deep-water periodic waves at {\em arbitrary} steepness. Remarkably, the model requires only 20 sets of HAM solutions to train an accurate predictive model that compute {\em convergent} series solution at {\em arbitrary} wave heights, including those approaching the limiting Stokes wave, within milliseconds. Compared to the relatively complex symbolic forms of traditional series solutions, the ML-based representations are significantly more compact and computationally efficient, while preserving comparable accuracy. 
Unlike other traditional methods that often focus solely on the surface profile or streamlines, our method recovers most essential physical quantities, including velocity fields, in the physical domain, enabling high-fidelity analysis of wave-induced flows in realistic oceanic environments.   

It should be emphasized that the proposed approach is not limited to wave problems mentioned in this paper. By using nonlinear water waves as a representative example, we establish a new {\em paradigm} to {\em quickly} obtain {\em convergent series} solutions  of complex nonlinear systems for a {\em whole} range of physical parameters,  thereby significantly broadening the scope of series solutions in science and engineering.

The remainder of this paper is organized as follows. Section~\ref{sec:governing_equations} presents the governing equations describing two-dimensional, steady, periodic Stokes waves in deep water, and outlines the construction of series solutions based on HAM. In Section~\ref{sec:results}, the HAM solutions are given and compared with classical solutions, followed by the development of a machine learning model capable of rapidly giving series Stokes wave solutions at infinite depth for arbitrary wave heights. Section~\ref{sec:inverse_mapping} describes a data-driven inverse mapping approach based on machine learning, enabling reconstruction of the wave solution in the physical plane. Finally, Section~\ref{sec:conclusion} summarizes the findings and outlines potential directions for future research.

\section{Mathematical description of Stokes wave in  infinite depth}\label{sec:governing_equations}

Consider a symmetric, two-dimensional, periodic gravity wave propagating from right to left with constant phase speed \( c \) over a seabed. The wave motion is affected only by gravity, and the bottom of the fluid domain is assumed to be horizontal. Since the phase speed \( c \) remains constant in an inertial reference frame, we consider another coordinate system moving with the wave. In this co-moving frame, the flow becomes steady.

The fluid is assumed to be inviscid, incompressible, and the motion to be irrotational, implying that the velocity field can be described by a complex analytic function of the variable \( z \). Therefore, the problem can be formulated using conformal mapping techniques.

\begin{figure}[t]
    \centering
    \begin{subfigure}{0.65\textwidth}
        \includegraphics[width=\linewidth]{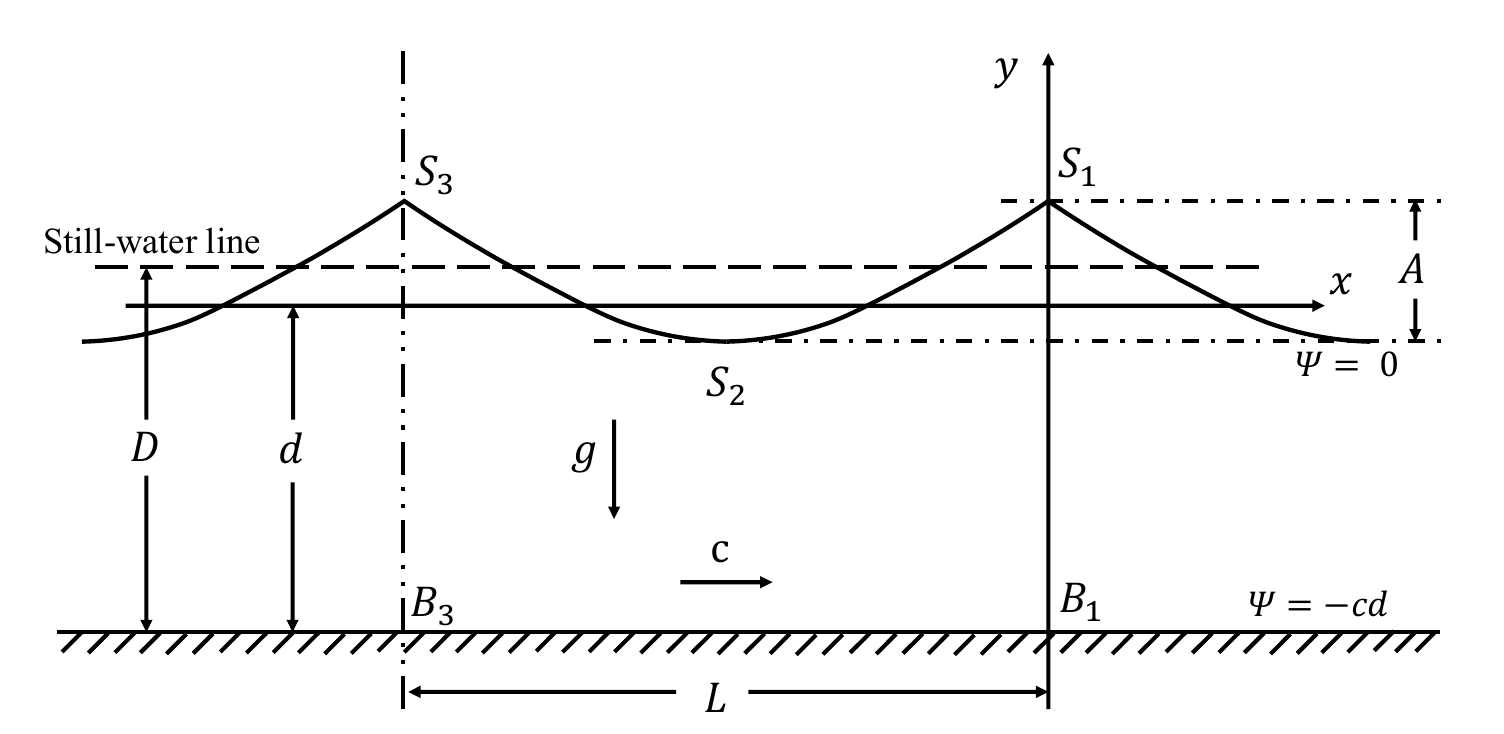}
        \caption{$z$ plane}
        \label{fig:z-plane}
    \end{subfigure}
    \begin{subfigure}{0.3\textwidth}
        \includegraphics[width=\linewidth]{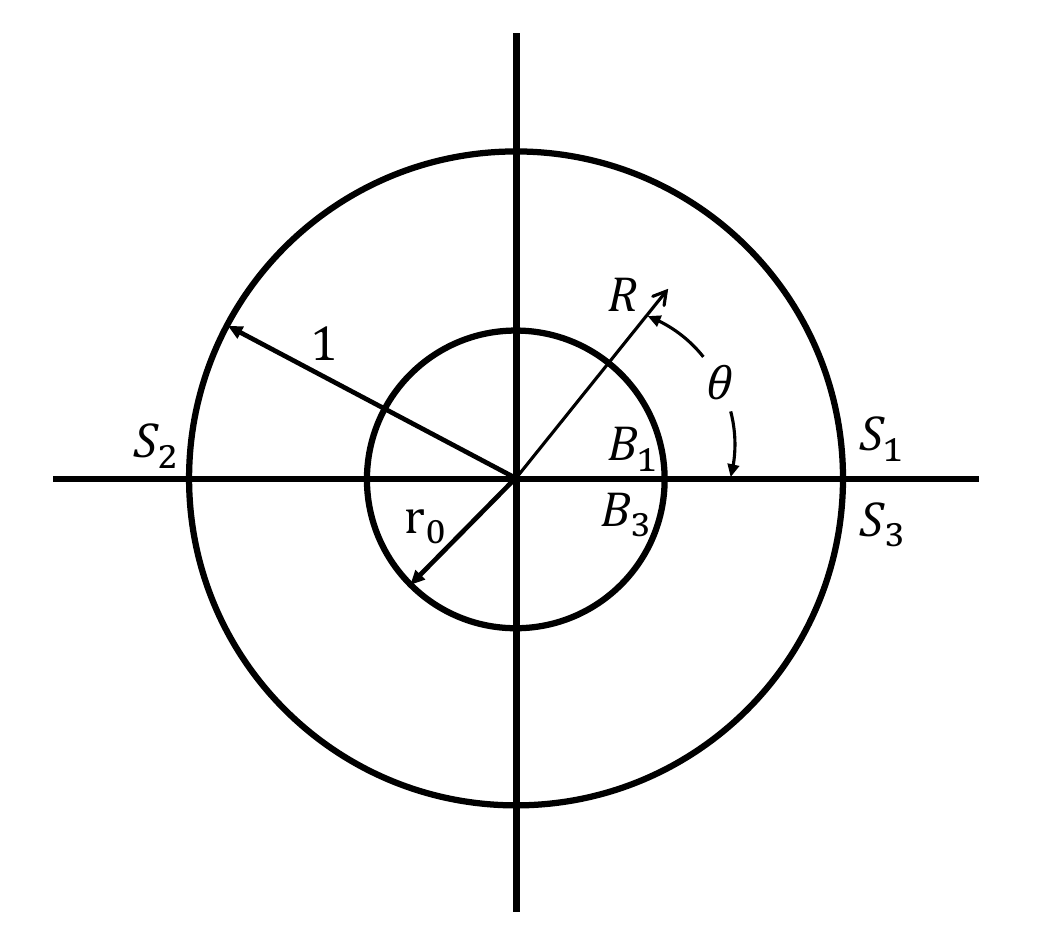}
        \caption{$\zeta$ plane}
        \label{fig:zeta-plane}
    \end{subfigure}
    \caption{(a) $z$ plane, (b) $\zeta$ plane.}
    \label{fig:plane}
\end{figure}

Figure~\ref{fig:z-plane} shows two periods of the wave in the physical plane (or \( z \) plane). In the figure, \( L \) denotes the wavelength, \( D \) is the mean water depth measured from the still water level, and \( A \) is the wave height, defined as the vertical distance between the wave crest and trough. The gravitational acceleration \( g \) acts vertically downward. Due to the symmetry of the wave profile, the vertical axis (y-axis) passes through a crest and serves as the axis of symmetry within one period. The horizontal axis (x-axis) is placed at a vertical distance \( d \) above the horizontal seabed. The mean elevation of the free surface is \(\bar y\) where an overbar denotes an average over one wave cycle. Therefore the mean depth is \(D = d + \bar y\) and does not in general equal \(d\). Since the fluid is assumed to be incompressible and irrotational, a velocity potential \(\Phi\) and a stream function \(\Psi\) may be defined, both of which satisfy the Laplace equation in the fluid domain:
\begin{equation}
\nabla \Phi^2 =0, \qquad \nabla \Psi^2= 0.
\label{eq:governing}
\end{equation}

The boundary conditions to be imposed on the flow are that the free surface and bottom are streamlines, that is:
\begin{subequations}\label{eq:psi}
\begin{align}
\Psi = 0 \qquad &\text{on} \qquad y = \eta, \\
\Psi = -cd \qquad &\text{on} \qquad y = -d,
\end{align}
\end{subequations}
where \(\eta\) denotes the free-surface elevation.

To nondimensionalize the equations, we choose a characteristic velocity scale \( c_0 = \sqrt{gL / 2\pi} \) and a characteristic length scale \( L / 2\pi \). All variables are then scaled using these reference quantities.

The Bernoulli condition on the free surface can be written in non-dimensional form as
\begin{equation}
|v|^{2} + 2y = K, \qquad \Psi = 0,
\label{eq:bernoulli-z}
\end{equation}
where the complex velocity is defined as \( v = v_x - i v_y \), \(|v|^{2} = v_x^{2}+v_y^{2}\) denotes the squared speed, and \( K \) is an unknown constant to be determined.

Since the unknown location of the free surface introduces substantial difficulty in solving the governing equations, we apply a conformal mapping from the physical \( z \) plane to an annular region in the \( \zeta \) plane, as illustrated in Figure~\ref{fig:zeta-plane}. This mapping is defined by the following transformation:
\begin{equation}
x+\mathrm{i} y=z(x, y)=z(\zeta)=\mathrm{i}\left[\ln \zeta+\sum_{j=1}^{+\infty} \frac{a_j}{j}\left(\zeta^j-\frac{r_0^{2 j}}{\zeta^j}\right)\right],
\label{eq:coordinate-trans}
\end{equation}
where $\zeta= Re^{i\theta}$, \( R \in [r_0, 1] \) is the radial coordinate, \( \theta \) is the angular coordinate, and \( r_0 \) denotes the inner radius of the annulus. The unknown coefficients \( a_1, a_2, \ldots, a_j, \ldots \) are constant real coefficients to be solved. Under this mapping, the horizontal bottom (streamfunction \( \Psi = -cd \)) is mapped to the inner circle \( R = r_0 = e^{-d} \), and the free surface (streamfunction \( \Psi = 0 \)) is mapped to the outer circle \( R = 1 \). In particular, wave crests correspond to the intersections of the outer circle with the positive $x$-axis at $\theta = 2\pi n$ (for example, \(S_1\) and \(S_3\)), and troughs correspond to the intersections with the negative $x$ axis at $\theta = (2n+1)\pi$ (for example, \(S_2\)). Furthermore, points labeled \(B\) denote the bed locations vertically below the corresponding surface points \(S\), in the mapped plane they lie on the inner circle $R = r_0$.

It is important to note that the limiting cases \( r_0 = 0 \) and \( r_0 = 1 \) correspond to infinite water depth and infinite shallow water, respectively. In this study, we focus on deep water and therefore set \(r_0=0\), which amounts to taking the still-water depth \(D\) to infinity.

The complex velocity potential \( w \) in the \( \zeta \) plane can be written in a simple form as:
\begin{equation}
w = \Phi + i \Psi = ic \ln \zeta = -c\theta + ic \ln R,
\label{eq:complex-velocity}
\end{equation}
where $\Phi$ represents the velocity potential.

Based on the mapping~\eqref{eq:coordinate-trans}, the physical coordinates can be expressed as a series expansion:
\begin{subequations}\label{eq:xy-Zeta}
\begin{align}
-x=\theta+\sum_{j=1}^{+\infty} \frac{a_j}{j}\left(R^j+\frac{r_0^{2 j}}{R^j}\right) \sin (j \theta), \\
y=\ln R+\sum_{j=1}^{+\infty} \frac{a_j}{j}\left(R^j-\frac{r_0^{2 j}}{R^j}\right) \cos (j \theta).
\end{align}
\end{subequations}

Therefore, the wavelength \( L \) of the wave is given by:
\begin{equation}
L =\left.x\right|_{R=1, \theta=0}-\left.x\right|_{R=1, \theta=2 \pi}=2 \pi,
\end{equation}
and wave steepness
\begin{equation}
\frac{A}{L}=\frac{1}{2 \pi}\left(\left.y\right|_{R=1, \theta=0}-\left.y\right|_{R=1, \theta=\pi}\right)=\sum_{j=1}^{+\infty} \frac{a_j}{2 j \pi}\left(1-r_0^{2 j}\right)[1-\cos (j \pi)].
\label{eq:steepness-equation}
\end{equation}

By substituting the transformed expressions into the Bernoulli equation and then equating the coefficients of $\cos\left(j\theta  \right)$, in combination with the wave steepness constraint ~\eqref{eq:steepness-equation} and the deep-water restriction \(r_0 = 0\), we obtain a new system of nonlinear algebraic equations.
\begin{subequations}\label{eq:Bernoulli-Zeta}
\begin{align}
    c^2 + 2 \sum_{l=1}^{\infty} & \frac{a_l f_l}{l} = K f_0, \label{eq:B-Z_a} \\
    \sum_{l=1}^{\infty} \frac{a_l}{l} \left\{ f_{|l-j|} + f_{j+l} \right\} &= K f_j \quad (j=1,2, \ldots), \label{eq:B-Z_b} \\
    \sum_{l=1}^{\infty} \frac{a_l}{l} (1 &- \cos (l \pi)) = A. \label{eq:B-Z_c}
\end{align}
\end{subequations}

For convenience, an auxiliary function \( f_i \), defined as:
\begin{subequations}\label{eq:f-series}
\begin{align}
    f_0 &= 1 + \sum_{l=1}^{\infty} a_l^2 , \label{eq:f0} \\
    f_1 &= a_1 + \sum_{l=1}^{\infty} a_l a_{l+1} , \label{eq:f1} \\
    f_j &= a_j + \sum_{l=1}^{\infty} a_l a_{l+j} \quad (j=2,3, \ldots) . \label{eq:fj}
\end{align}
\end{subequations}

By truncating the series expansion to order \( r \), we obtain a total of \( r + 2 \) equations. Among these, \( r \) equations are independent of the wave speed \( c \) and are written as:
\begin{equation}
    \mathcal{N}_j(K, a_1, a_2, \ldots, a_r) = \sum_{l=1}^{r} \frac{a_l}{l}\left\{f_{|l-j|}+f_{j+l}\right\} - K f_j = 0 \quad(j=1,2, \ldots, r).
    \label{eq:nonlinear_eq}
\end{equation}

This system~\eqref{eq:nonlinear_eq} and ~\eqref{eq:B-Z_c} can be solved for the \( r + 1 \) unknown Fourier coefficients \( a_1, a_2, \ldots, a_r \) and the Bernoulli constant \( K \).

Next, the homotopy analysis method is used to solve the nonlinear equations~\eqref{eq:nonlinear_eq}. Let \( a_{j,0} \) denote the initial guess of the unknown coefficient \( a_j \) (\( j = 1, 2, \ldots, r \)), \(K_0\) denote the initial guess of the unknown constant \( K \), and let \( \hbar \neq 0 \) be an auxiliary parameter (called the convergence-control parameter). Introduce the embedding parameter \( q \in [0,1] \) and construct a family of equations in the form:
\begin{subequations}\label{eq:HAM_0}
\begin{align}
    (1-q)[K(q)-K_0] &= \hbar q \mathcal{N}_1[K(q),\Omega_1(q), \Omega_2(q),\ldots,\Omega_r(q)], \label{eq:HAM_0a} \\
    (1-q)[\Omega_{j-1}(q)-a_{j-1,0}] &= \hbar q \mathcal{N}_j[K(q),\Omega_1(q), \Omega_2(q),\ldots,\Omega_r(q)] \quad (j=2,3, \ldots,r). \label{eq:HAM_0b}
\end{align}
\end{subequations}

The nonlinear operators $\mathcal{N}_1, \mathcal{N}_2, \ldots, \mathcal{N}_r$ are defined by equation~\eqref{eq:nonlinear_eq}, and the unknown functions $K(q)$,  $\Omega_1(q)$, $\Omega_2(q)$, $\ldots$, $\Omega_r(q)$ represent the continuously deformed approximations of the unknown constant $K$ and Fourier coefficients $a_1, a_2, \ldots, a_r$, respectively. The initial guesses are denoted as $a_{j,0}$, where $a_{j,0}$ serves as an initial approximation of $a_j$. The convergence-control parameter $\hbar$, offers a simple and effective way to ensure the convergence of the solution series. If \( \hbar \) is chosen such that the series converges at \( q = 1 \), then the homotopy-series solutions of the original problem is given by: (detailed derivation is shown in appendix~\ref{appA})
\begin{subequations}
\begin{align}
K &= \sum_{k=0}^{\infty}K_{k}, \\
a_j &= \sum_{k=0}^{\infty}a_{j,k} \quad (j=1,2,3,\ldots,r).
\end{align}
\end{subequations}

The \( n \)-th order homotopy approximation of the solution is then written as:
\begin{subequations}\label{eq:nth_order_ham_solution}
\begin{align}
\tilde{K}_n &= \sum_{k=0}^{n}K_{k}, \\
\tilde{\Omega}_{j,n} &= \sum_{k=0}^{n}a_{j,k} \quad (j=1,2,3,\ldots,r). 
\end{align}
\end{subequations}

After determining all the Fourier coefficients \( a_1, a_2, \ldots, a_r \) and the parameter \( K \), the wave speed \( c \) can be directly evaluated using equation~\eqref{eq:get_c}:
\begin{equation}
    c^2 = - 2 \sum_{l=1}^{r} \frac{a_l f_l}{l} + K f_0.
    \label{eq:get_c}
\end{equation}

To quantify the global error of the HAM approximation, we define the total squared residual error as follows:
\begin{equation}
    \varepsilon = \sum_{j=1}^{r} \left( \mathcal{N}_j[\tilde{K},\tilde{\Omega_1}, \tilde{\Omega_2},\ldots,\tilde{\Omega_r}] \right)^2,
    \label{eq:HAM_var}
\end{equation}
where \( \mathcal{N}_j \) are the nonlinear operators from~\eqref{eq:nonlinear_eq}. Obviously, the smaller the residual \( \varepsilon \), the more accurate the homotopy approximation~\eqref{eq:nth_order_ham_solution}. According to \citet{liao2003beyond, liao2012ham}, if all squared residuals approach zero, then the homotopy series converges to the exact solution of the original nonlinear problem.Therefore, it is sufficient to test the squared residuals~\eqref{eq:HAM_var}.

\section{Solutions in \(\zeta\) plane}\label{sec:results}
\subsection{HAM solutions for discrete physical parameters}\label{sec:HAM-results}

Based on the above methodology, we are able to obtain a convergent homotopy series solution for Stokes waves of arbitrary wave height of infinite water depth. 

\begin{figure}[t]
    \centering
    \includegraphics[width=0.45\textwidth, trim=10pt 10pt 10pt 10pt, clip]{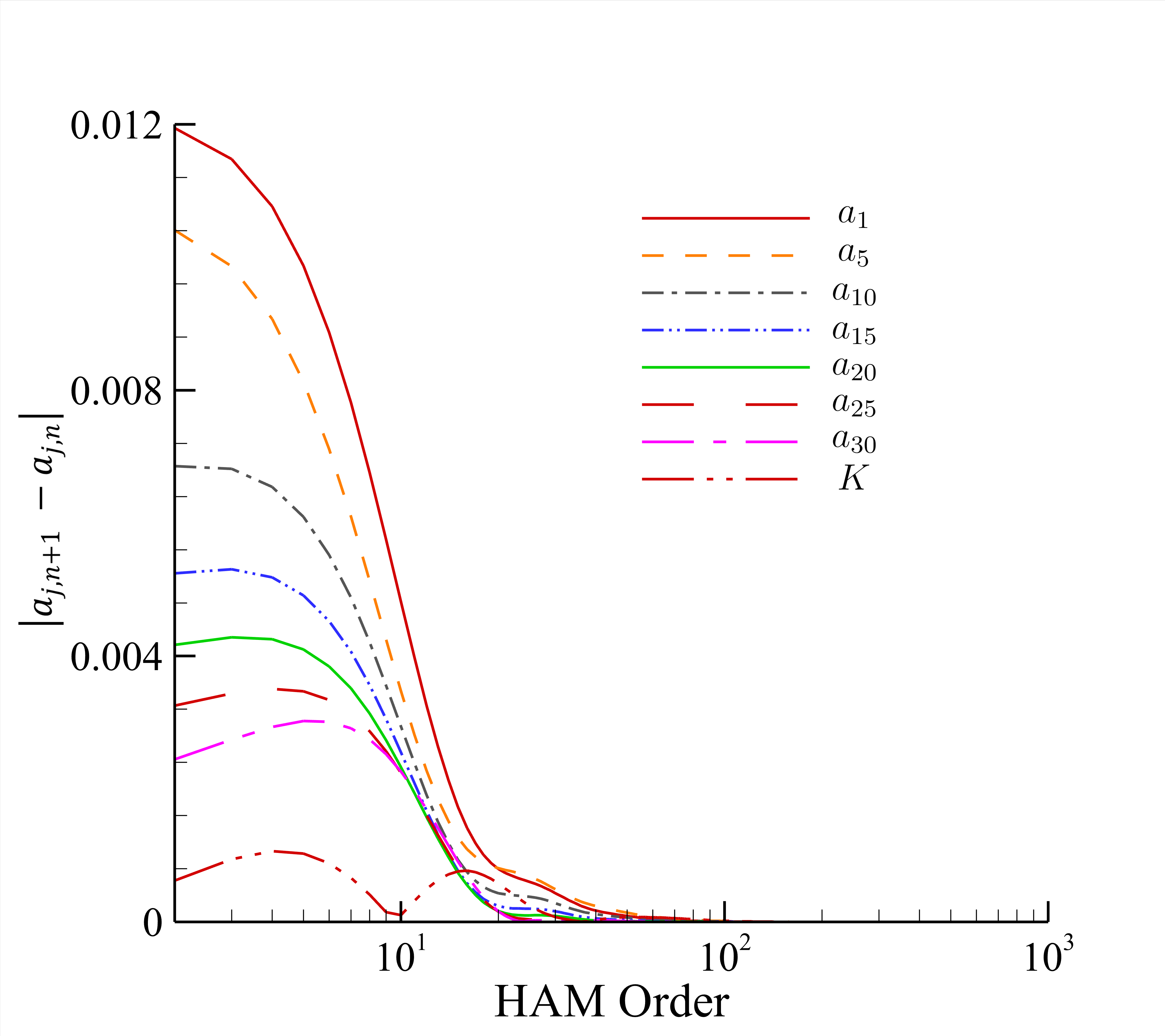}
    \caption{Convergence behavior of selected Fourier coefficients \( a_1 \), \( a_5 \), \( a_{10} \), \( a_{15} \), \( a_{20} \), \( a_{25} \), \( a_{30} \), and the Bernoulli constant \( K \) for a Stokes wave in infinite depth with steepness \( A/L = 0.07 \), based on 1000th-order HAM approximations. The y-axis is the successive order difference $|a_{j,n+1}-a_{j,n}|$ for coefficient $a_j$, and the x-axis is the HAM order $n$.}
    \label{fig:aj_convergency}
\end{figure}

To evaluate the convergence characteristics of the HAM series solution, we consider a wave steepness of \( A/L = 0.07 \) and retain 30 Fourier coefficients \(a_j\) in the expansion. The homotopy approximations are computed up to the 1000th order. The convergence behavior of the selected Fourier coefficients \( a_1 \), \( a_5 \), \( a_{10} \), \( a_{15} \), \( a_{20} \), \( a_{25} \), \( a_{30} \), and the Bernoulli constant \( K \) is illustrated in Figure~\ref{fig:aj_convergency}. The horizontal axis represents the order of the homotopy series solution on a logarithmic scale, while the vertical axis shows the absolute difference between successive approximations of the corresponding physical quantities. A decreasing trend towards zero indicates convergence of the series. As shown in the figure, the differences diminish rapidly with increasing order, demonstrating that all coefficients converge swiftly and with high precision.

\begin{figure}[t]
    \centering
    \includegraphics[width=0.45\textwidth, trim=10pt 10pt 10pt 10pt, clip]{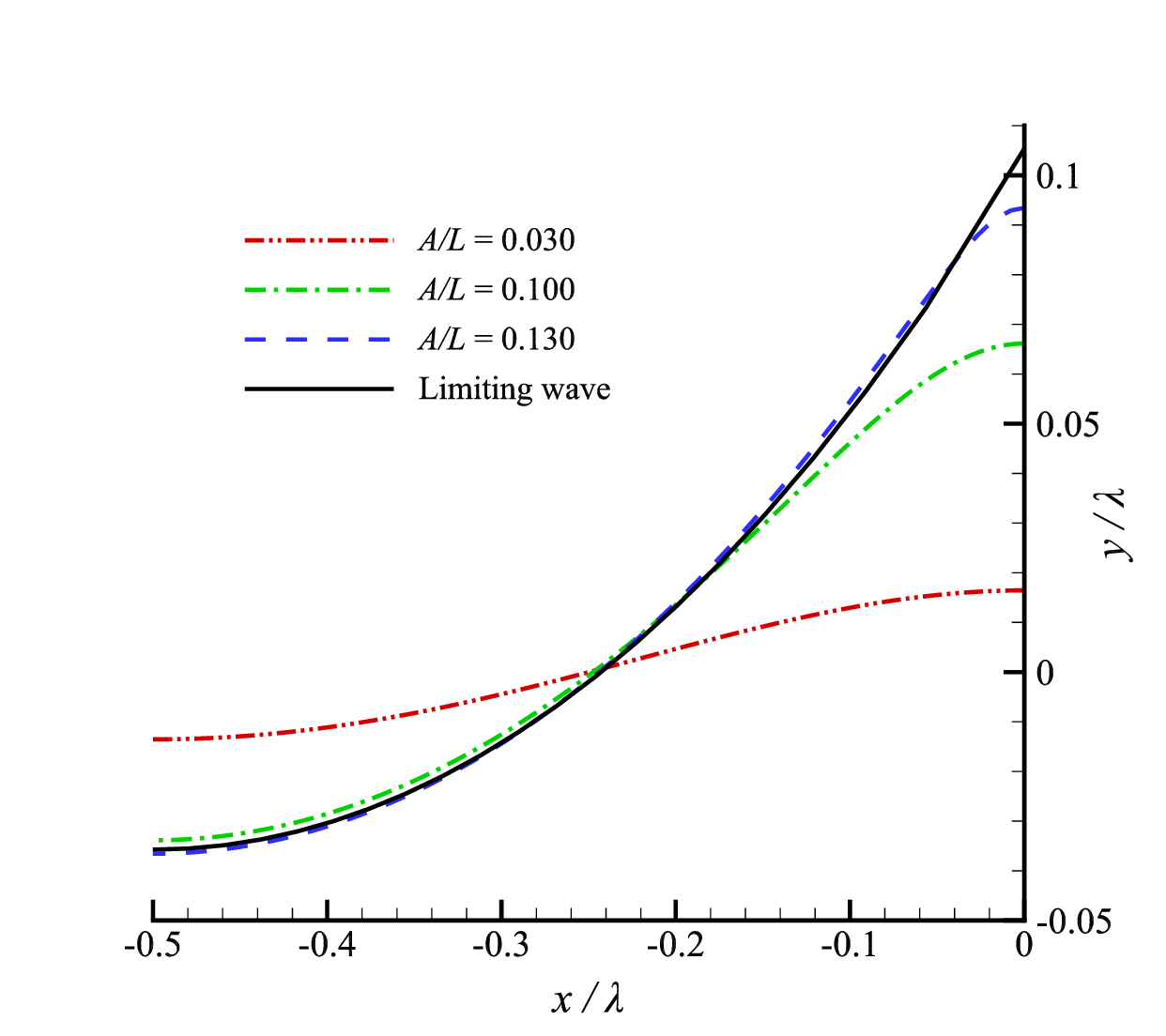}
    \caption{Stokes wave profiles in infinite depth for different wave steepness values.}
    \label{fig:different_steepness}
\end{figure}

Figure~\ref{fig:different_steepness} shows the Stokes wave solutions at infinite water depth for different steepness values, illustrating the effect of wave steepness on the Stokes wave profile. As anticipated, increasing wave steepness significantly intensifies the nonlinear features of the free surface. Notably, the wave crest becomes increasingly acute, tending toward a pointed tip, while the trough broadens and flattens, leading to a pronounced deviation from sinusoidal symmetry. Of particular significance is the limiting case where the wave steepness approaches \( A/L = 0.141 \). In this case, our calculations reveal that the crest angle sharpens to approximately \(119.48^\circ\), closely matching the classical theoretical prediction of \(120^\circ\) for Stokes' highest wave. This quantitative agreement provides strong validation of the HAM in capturing extreme nonlinear wave dynamics without relying on numerical extrapolation or empirical correction. It is also worth emphasizing that as the wave steepness increases, the associated wave profile becomes increasingly rich in high-frequency components. To accurately capture such sharply peaked profiles, a higher number of Fourier modes is required. In our formulation, the number of retained Fourier components is increased to ensure convergence and numerical precision across varying steepness values.

To quantitatively validate the accuracy of the present method, the squared wave velocity \( c^2 \) and the Bernoulli constant \( K \) calculated for selected wave steepness values in infinite depth are listed in Table~\ref{tab:comparison_with_Schwartz}, alongside the data reported by~\citet{schwartzComputerExtensionAnalytic1974a}. The results demonstrate excellent agreement across all cases, affirming both the reliability and physical consistency of our method. In particular, for the limiting Stokes wave, our method is also able to guarantee both convergence and high accuracy, further highlighting the efficacy of the HAM formulation in modeling strongly nonlinear free-surface waves.

\begin{table}
    \centering
    \begin{tabular}{ccccc}
        \toprule
        \multirow{2}{*}{\textit{A/L}} & \multicolumn{2}{c}{$c^2$} & \multicolumn{2}{c}{$K$} \\
         \cmidrule(lr){2-3} \cmidrule(lr){4-5}
        & \citet{schwartzComputerExtensionAnalytic1974a} & HAM & \citet{schwartzComputerExtensionAnalytic1974a} & HAM \\
        \midrule
        0.040 & 1.01592 & 1.019107& 1.03145& 1.034538\\
        0.070 &1.04955 & 1.049952& 1.09533& 1.095726\\
        0.100 & 1.10367& 1.103807& 1.19111& 1.191081 \\
        0.120 & 1.15182& 1.151767& 1.26790& 1.267701\\
        0.130 & 1.17820& 1.177986&1.30546& 1.305137\\
        0.135 &1.18996 & 1.189858& 1.32017& 1.320065\\
        0.140 &1.193 & 1.195776&1.3207 & 1.324652\\
        limiting wave & - & 1.193987& -& 1.321443\\
        \bottomrule
    \end{tabular}
    \caption{Comparison of the squared wave velocity \( c^2 \) and the Bernoulli constant \( K \) for Stokes waves of various steepness values in infinite depth water between the present HAM-based solutions and the reference data reported by~\citet{schwartzComputerExtensionAnalytic1974a}.}
    \label{tab:comparison_with_Schwartz}
\end{table}

\begin{figure}[t]  
    \centering
    \includegraphics[width=0.5\textwidth, trim=10pt 10pt 10pt 10pt, clip]{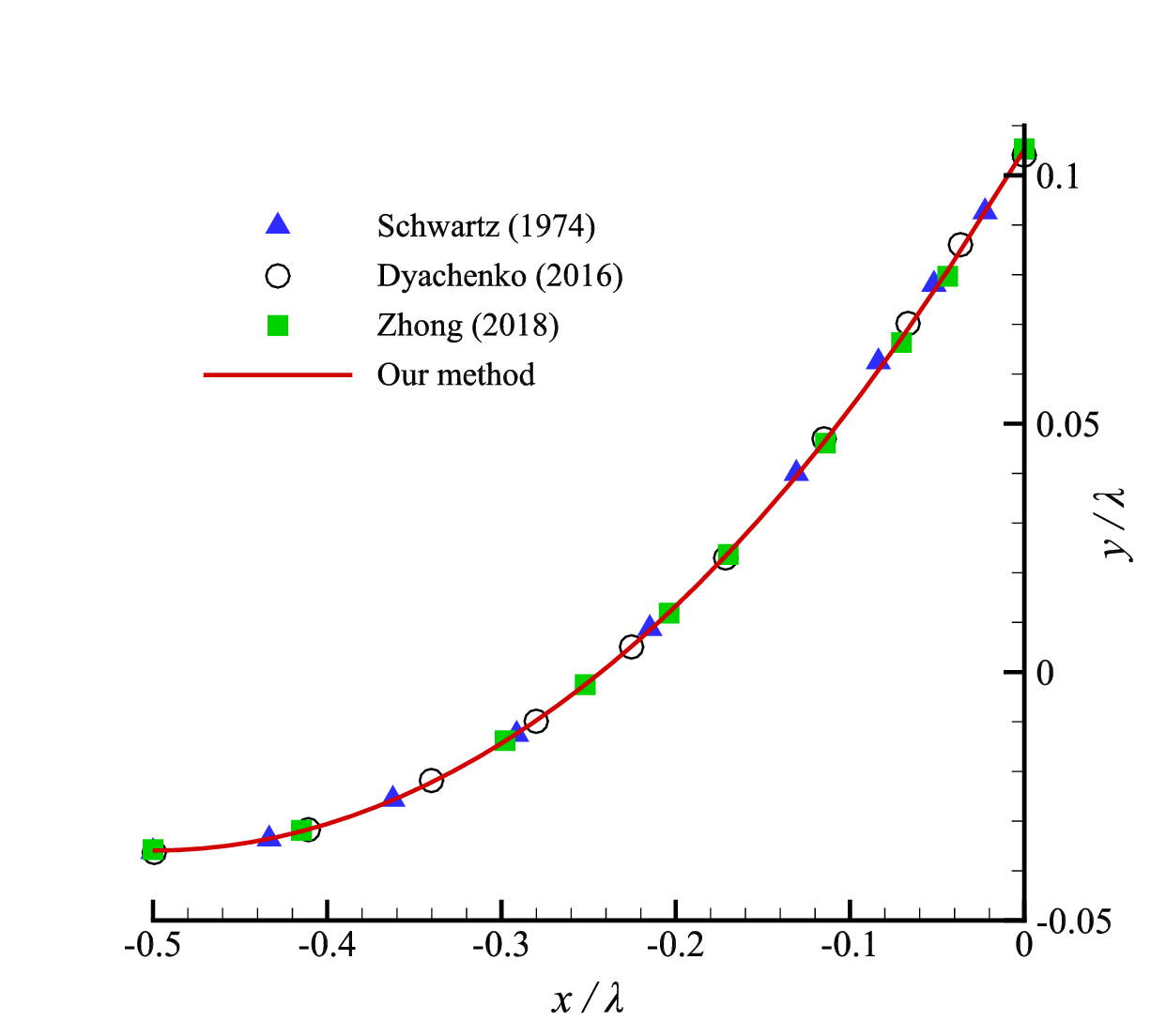}
    \caption{Comparison of limiting Stokes wave profiles in infinite depth water obtained by different methods.}
    \label{fig:comparison_limit}
\end{figure}

Figure~\ref{fig:comparison_limit} presents a comparative analysis of the limiting Stokes wave profiles obtained by several different approaches, including those of ~\citet{schwartzComputerExtensionAnalytic1974a}, \citet{dyachenko2016branch}, \citet{zhong2018limiting}, and our present method. Notably, the wave profile produced by our HAM solution exhibits excellent agreement with the benchmark results reported in the literature. 

Furthermore, based on the converged high-order Fourier coefficients obtained from our method, we can reconstruct the complete velocity potential and thus recover detailed information about the underlying flow field. In principle, this allows for the calculation of key flow characteristics such as velocity components, and pressure distributions throughout the fluid domain. 

\begin{figure}[t]
    \centering
    \begin{subfigure}{0.42\textwidth}
        \includegraphics[width=\linewidth, trim=10pt 10pt 10pt 10pt, clip]{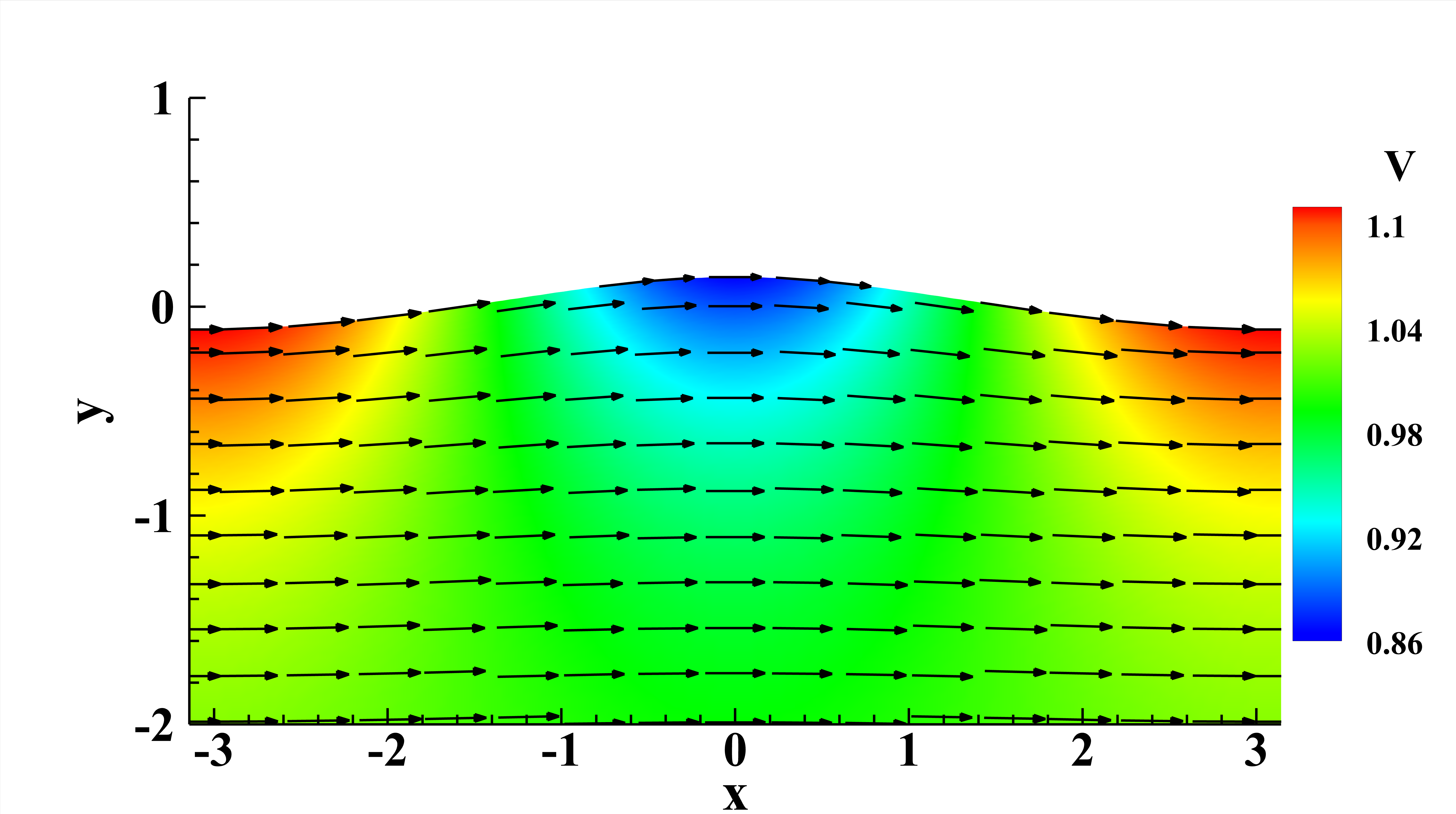}
        \caption{$A/L=0.040$}
        \label{HAM-0.040}
    \end{subfigure}
    \begin{subfigure}{0.42\textwidth}
        \includegraphics[width=\linewidth, trim=10pt 10pt 10pt 10pt, clip]{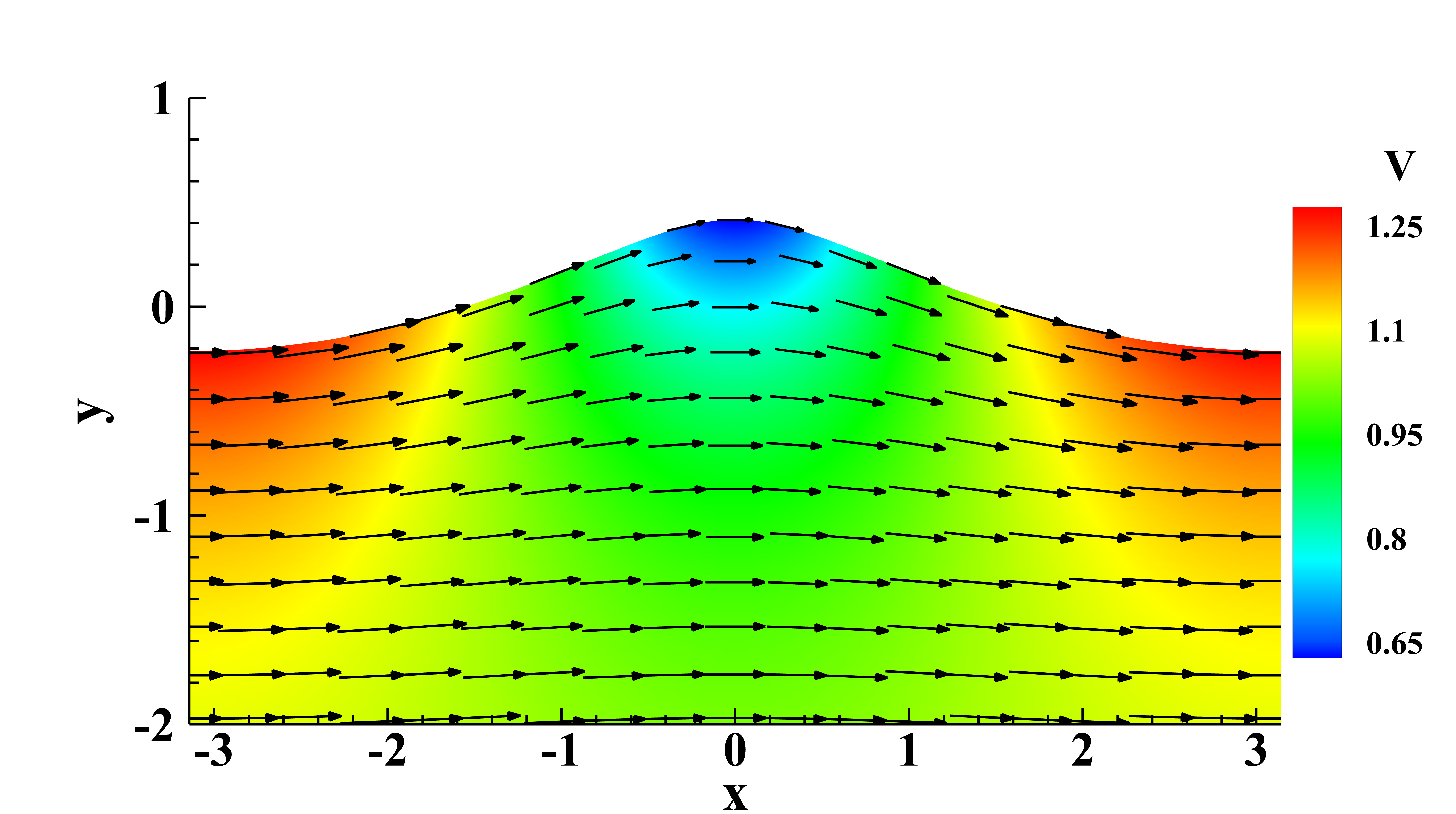}
        \caption{$A/L=0.100$}
        \label{HAM-0.100}
    \end{subfigure}
    \begin{subfigure}{0.42\textwidth}
        \includegraphics[width=\linewidth, trim=10pt 10pt 10pt 10pt, clip]{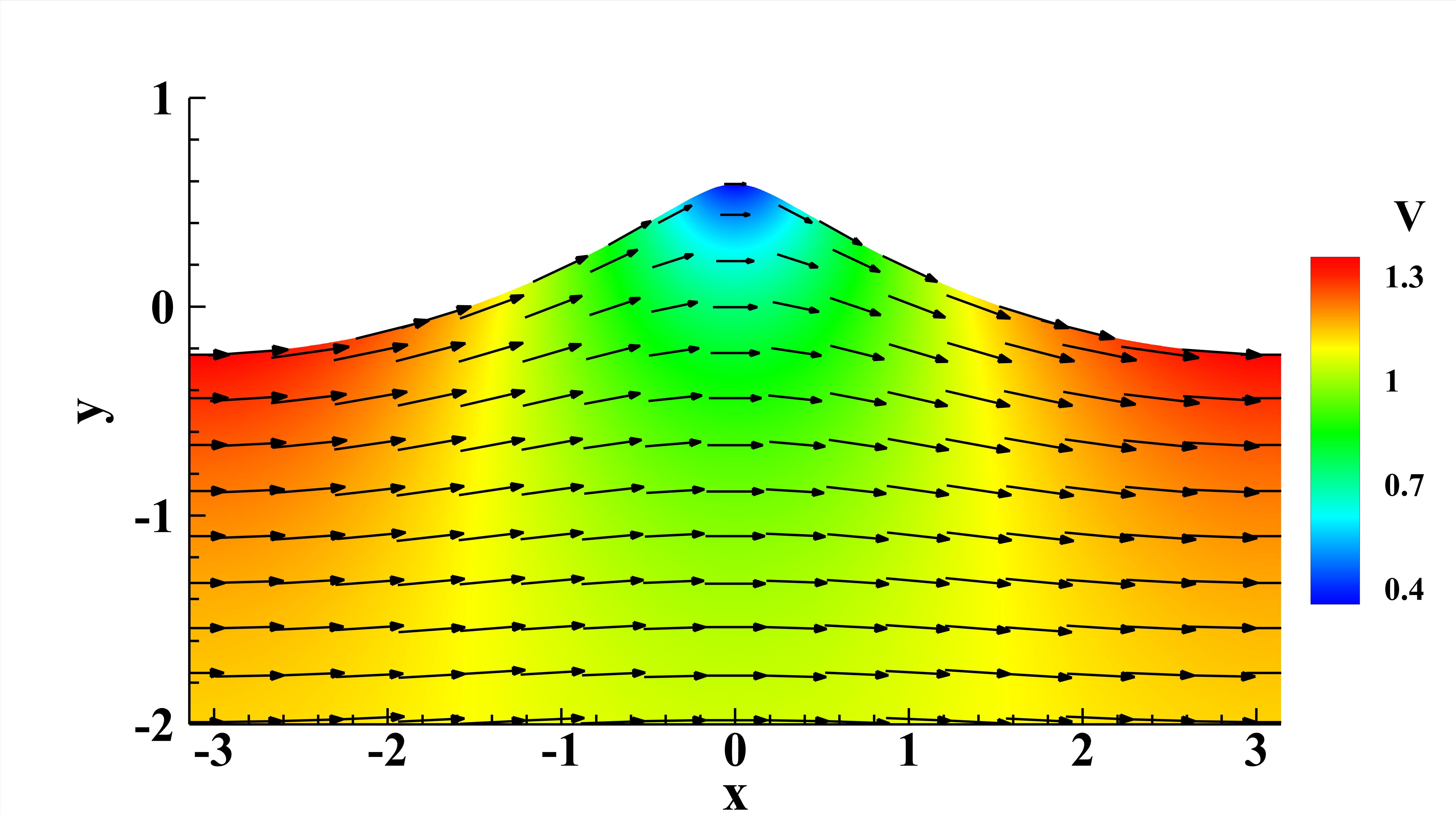}
        \caption{$A/L=0.130$}
        \label{HAM-0.130}
    \end{subfigure}
    \begin{subfigure}{0.42\textwidth}
        \includegraphics[width=\linewidth, trim=10pt 10pt 10pt 10pt, clip]{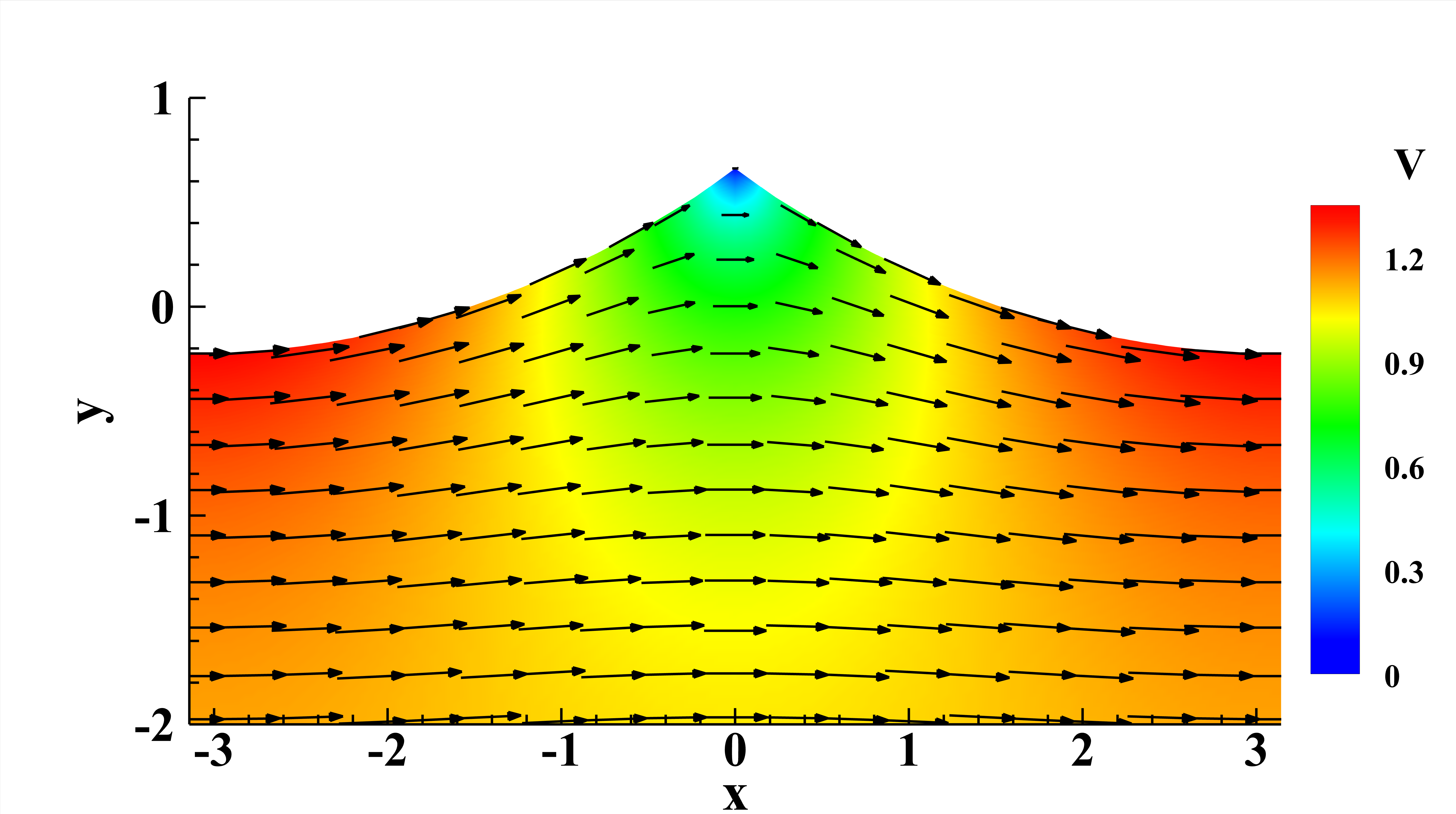}
        \caption{Limiting wave}
        \label{HAM-0.141}
    \end{subfigure}
    \caption{Stokes wave flow fields of varying steepness in infinite-depth water. Arrows represent the local velocity vectors, while the color contours depict the magnitude of the flow speed.}
    \label{fig:HAM-profile-comparison}
\end{figure}

Figure~\ref{fig:HAM-profile-comparison} illustrates the velocity fields corresponding to several representative wave profiles with different steepness values. In these visualizations, the arrows denote the local flow directions (velocity vectors), while the color contours indicate the magnitude of the resultant velocity. As expected, steeper waves exhibit more pronounced velocity gradients beneath the crest, reflecting the intensification of nonlinear interactions within the fluid bulk.

For two-dimensional Stokes waves of arbitrary wave height in infinite-depth water, the proposed HAM-based framework enables the construction of series solutions in a completely self-contained manner. Unlike traditional perturbation or numerical methods, our approach does not require any small parameters, linearization assumptions, or external convergence acceleration techniques. Instead, the HAM offers a flexible mechanism for directly generating convergent series solutions, where the accuracy can be systematically improved by increasing the number of retained Fourier coefficients and the order of the homotopy expansion.

\subsection{ML solutions for arbitrary physical parameters}\label{sec:ml-results}
From the analysis presented in the previous section, it is evident that the HAM can be employed to construct series solutions for two-dimensional Stokes waves of arbitrary wave height in infinite depth water. Through this framework, we can directly obtain a wide range of essential wave characteristics, including the wave profile, phase speed, and internal velocity fields, with high accuracy.

However, a notable limitation of the classical HAM lies in its computational cost: for each new value of wave steepness, the entire process (including the construction of nonlinear algebraic systems, convergence control, and high-order series calculation) must be recalculated from scratch. This leads to considerable computational overhead. Such inefficiency presents a major bottleneck when applying HAM to practical engineering problems or large-scale simulations.

This naturally raises the following question: is it possible to achieve rapid and accurate prediction of nonlinear wave solutions across a continuous range of wave steepness values, without the need for repeated full-order HAM calculations? With the advent of modern machine learning (ML) techniques, particularly deep learning models designed for scientific computing, this challenge can now be addressed from a new perspective. Data-driven machine learning models, trained on a set of representative HAM solutions, have the potential to approximate the complex function between wave steepness and wave features with high fidelity, while drastically reducing the computational time. In the following sections, we explore a hybrid approach that leverages the analytical rigor of HAM and the predictive efficiency of machine learning. This strategy not only preserves the theoretical accuracy of the original method, but also enables real-time, scalable prediction of strongly nonlinear wave phenomena.

\begin{figure}[t]
    \centering
    \includegraphics[width=0.65\textwidth]{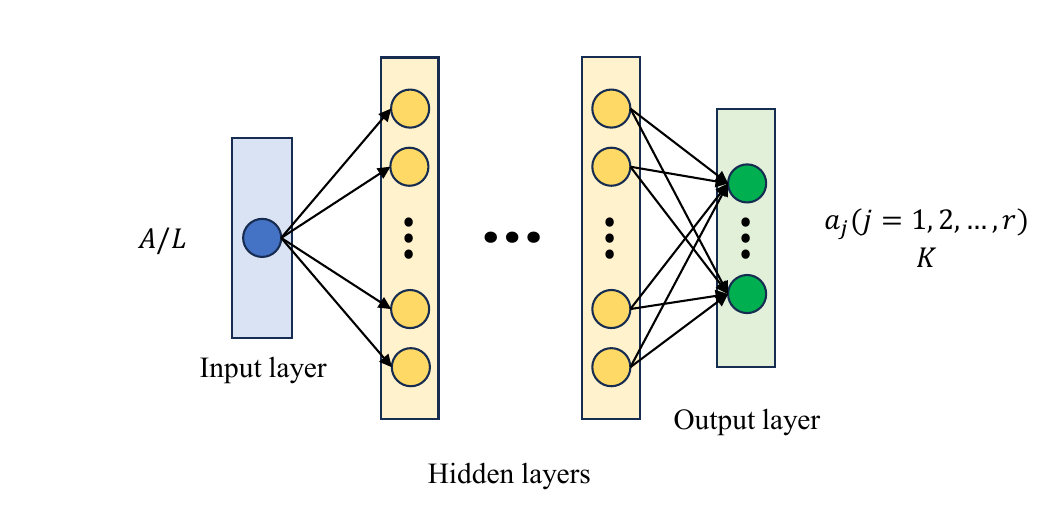}
    \caption{Schematic of machine learning model used to predict the solution of the wave in infinite depth water with arbitrary wave height.  The corresponding ML codes can be free downloaded on the website  \url{https://github.com/sjtu-liao/deep-water-Stokes-wave} and \url{https://numericaltank.sjtu.edu.cn/NonlinearWaves.htm} (see ``Convergent Series of Stokes Wave of Arbitrary Height in Deep Water'').}
    \label{fig:DNN-1}
\end{figure}

We construct a fully connected deep artificial neural network (DNN) architecture in which the input is the wave steepness, and the output is a high-dimensional vector containing the Fourier coefficients $a_j$ and Bernoulli constant $K$, which together constitute the solution of the Stokes wave as established in the previous section. The schematic of the DNN model is shown in Figure~\ref{fig:DNN-1}.   The corresponding ML codes can be free downloaded on the website  \url{https://github.com/sjtu-liao/deep-water-Stokes-wave} and \url{https://numericaltank.sjtu.edu.cn/NonlinearWaves.htm} (see ``Convergent Series of Stokes Wave of Arbitrary Height in Deep Water'').

Prior to network training, we systematically analyzed the required number of Fourier modes necessary to accurately represent Stokes waves of different steepness values. Waves with small steepness exhibit weak nonlinearity and can be adequately captured using a small number of Fourier modes. In contrast, waves with high steepness, especially near the limiting case, display strong nonlinear characteristics, including sharp crests and flattened troughs, which necessitate a significantly larger number of Fourier modes for accurate reconstruction. This modal analysis ensures that the neural network is trained on representations with sufficient resolution across the full steepness range. Based on the above analysis, we used the HAM to gain accurate series solutions for $20$ representative wave steepness values ranging from weakly to strongly nonlinear regimes. These solutions, containing their respective Fourier coefficients $a_j$ and Bernoulli constant $K$, were compiled into the training dataset. The input-output pairs were then used to train the neural network using a mean squared error (MSE) loss function and an adaptive gradient-based optimizer.

By training the DNN model on the carefully curated dataset of wave solutions computed via the HAM, we effectively establish a continuous nonlinear mapping from the wave steepness to the complete wave field representation. Once trained, the DNN model is capable of providing accurate predictions of the wave profile and related flow quantities for any given wave steepness, typically within milliseconds. This enables near real-time evaluation of wave parameters, which is highly advantageous for large-scale parametric studies and engineering applications such as wave forecasting and wave-structure interaction modeling.

\begin{figure}[t]
    \centering
    \begin{subfigure}{0.3\textwidth}
        \includegraphics[width=\linewidth, trim=10pt 10pt 10pt 10pt, clip]{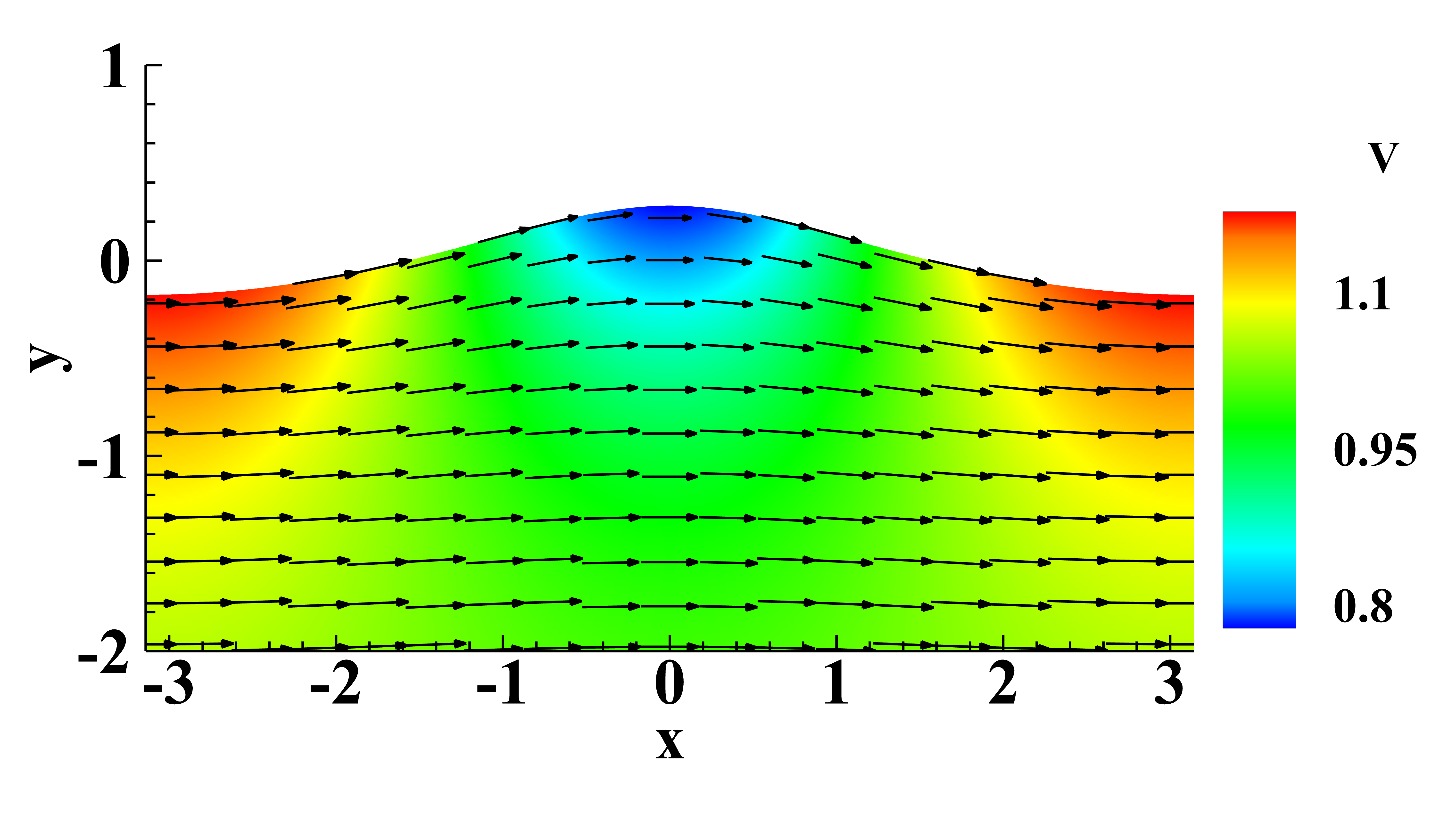}
        \caption{ML, $A/L=0.070$}
    \end{subfigure}
    \begin{subfigure}{0.3\textwidth}
        \includegraphics[width=\linewidth, trim=10pt 10pt 10pt 10pt, clip]{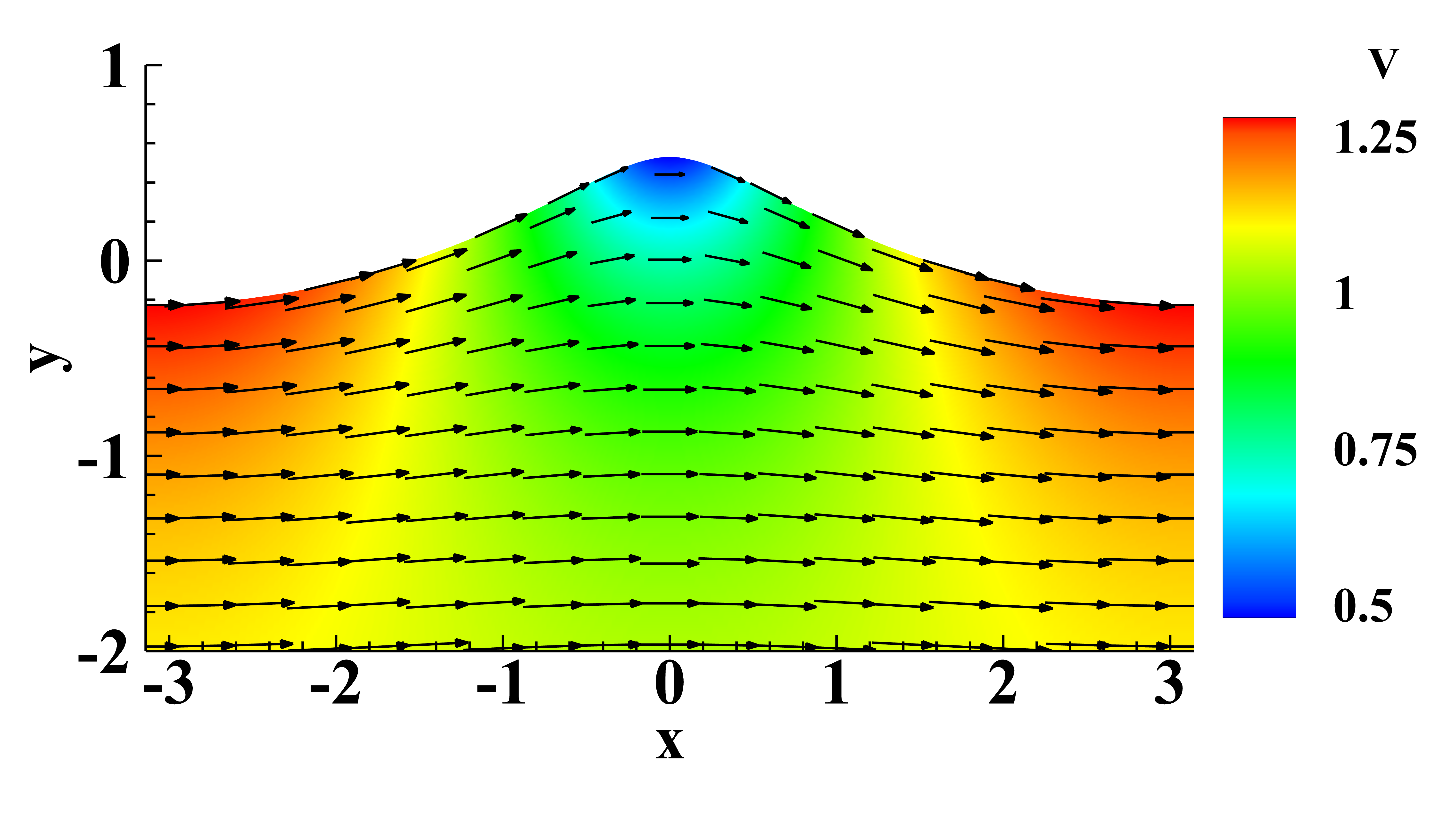}
        \caption{ML, $A/L=0.120$}
    \end{subfigure}
    \begin{subfigure}{0.3\textwidth}
        \includegraphics[width=\linewidth, trim=10pt 10pt 10pt 10pt, clip]{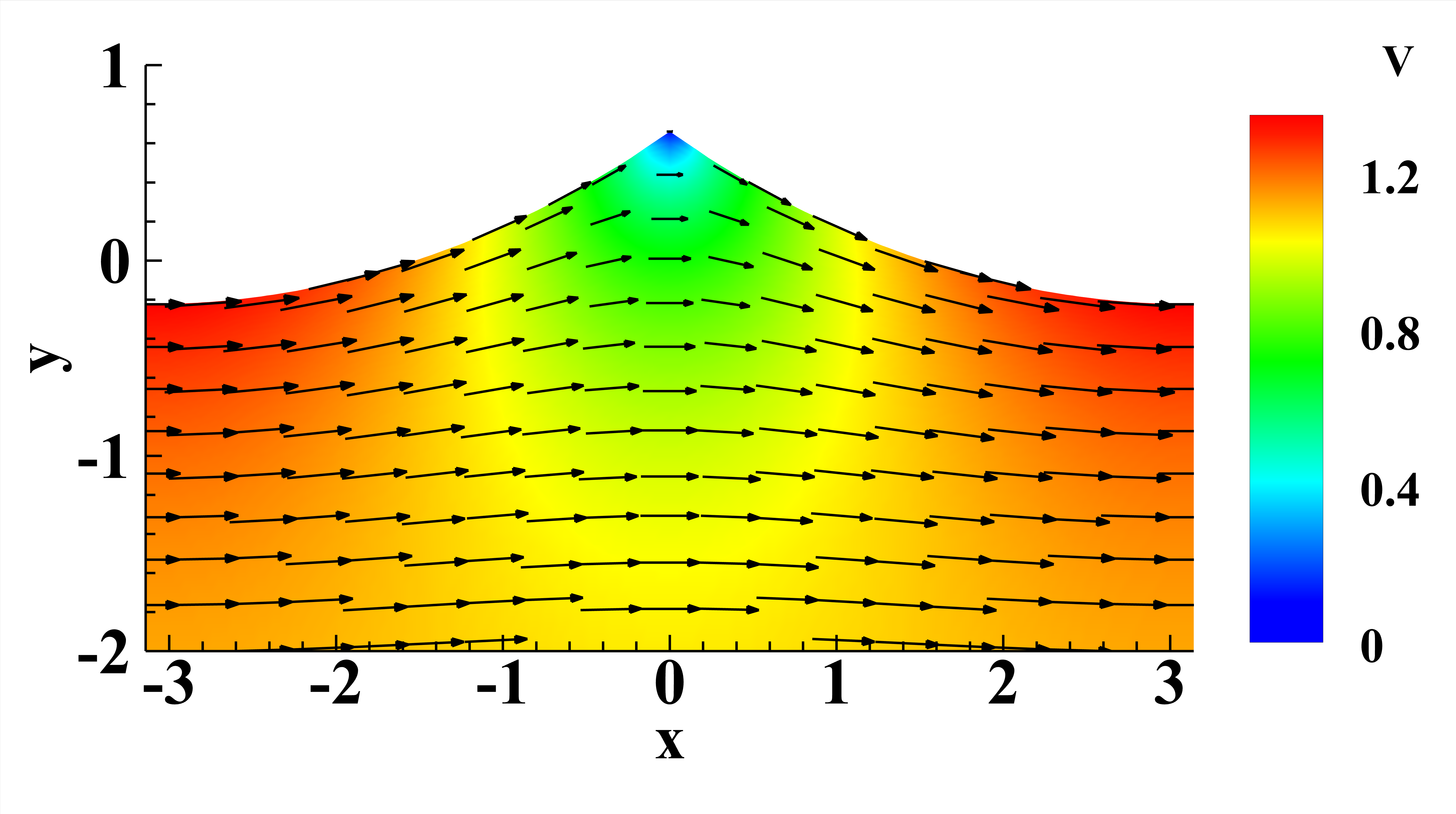}
        \caption{ML, Limiting wave}
    \end{subfigure}
    
    \begin{subfigure}{0.3\textwidth}
        \includegraphics[width=\linewidth, trim=10pt 10pt 10pt 10pt, clip]{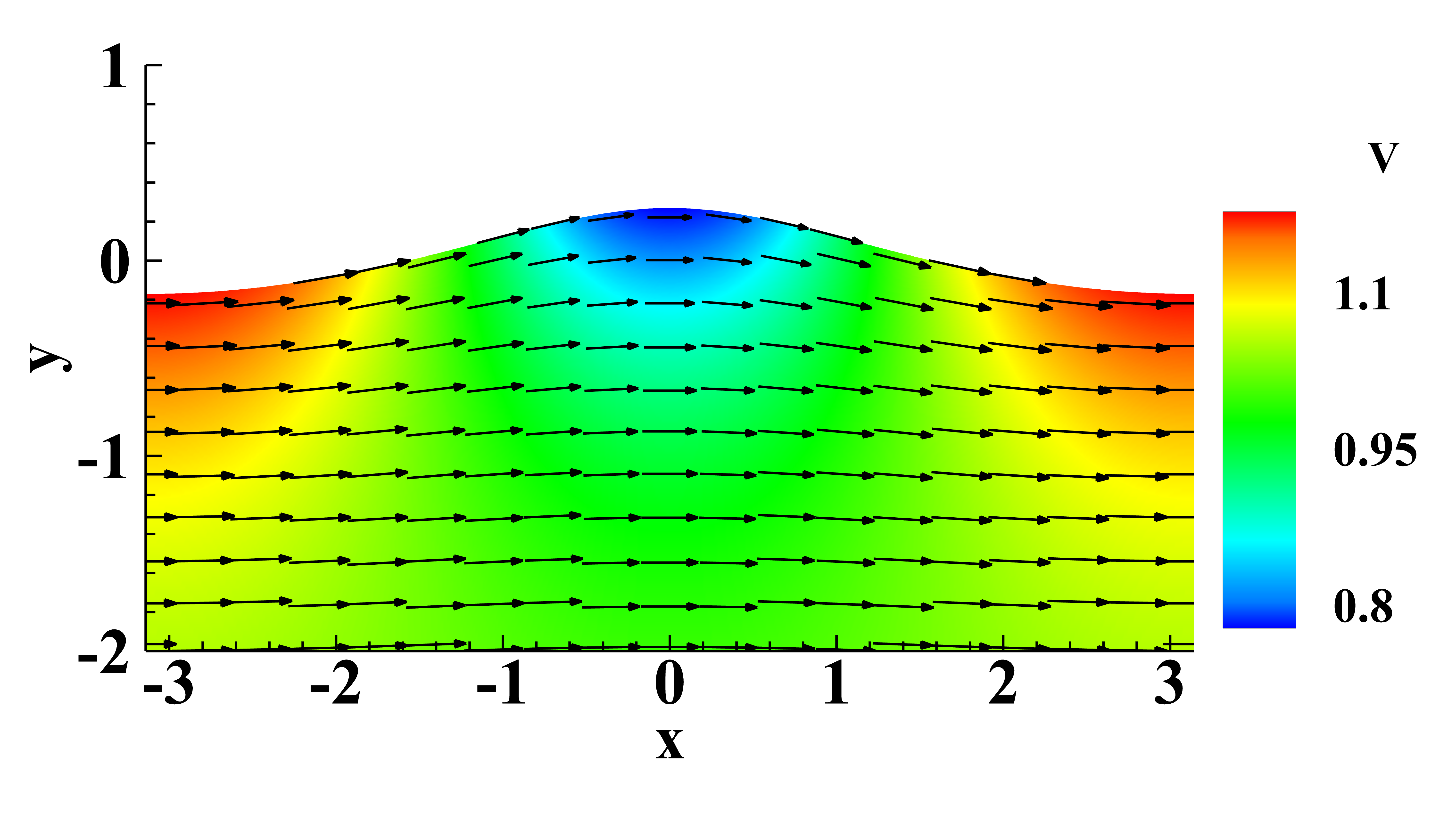}
        \caption{HAM, $A/L=0.070$}
    \end{subfigure}
    \begin{subfigure}{0.3\textwidth}
        \includegraphics[width=\linewidth, trim=10pt 10pt 10pt 10pt, clip]{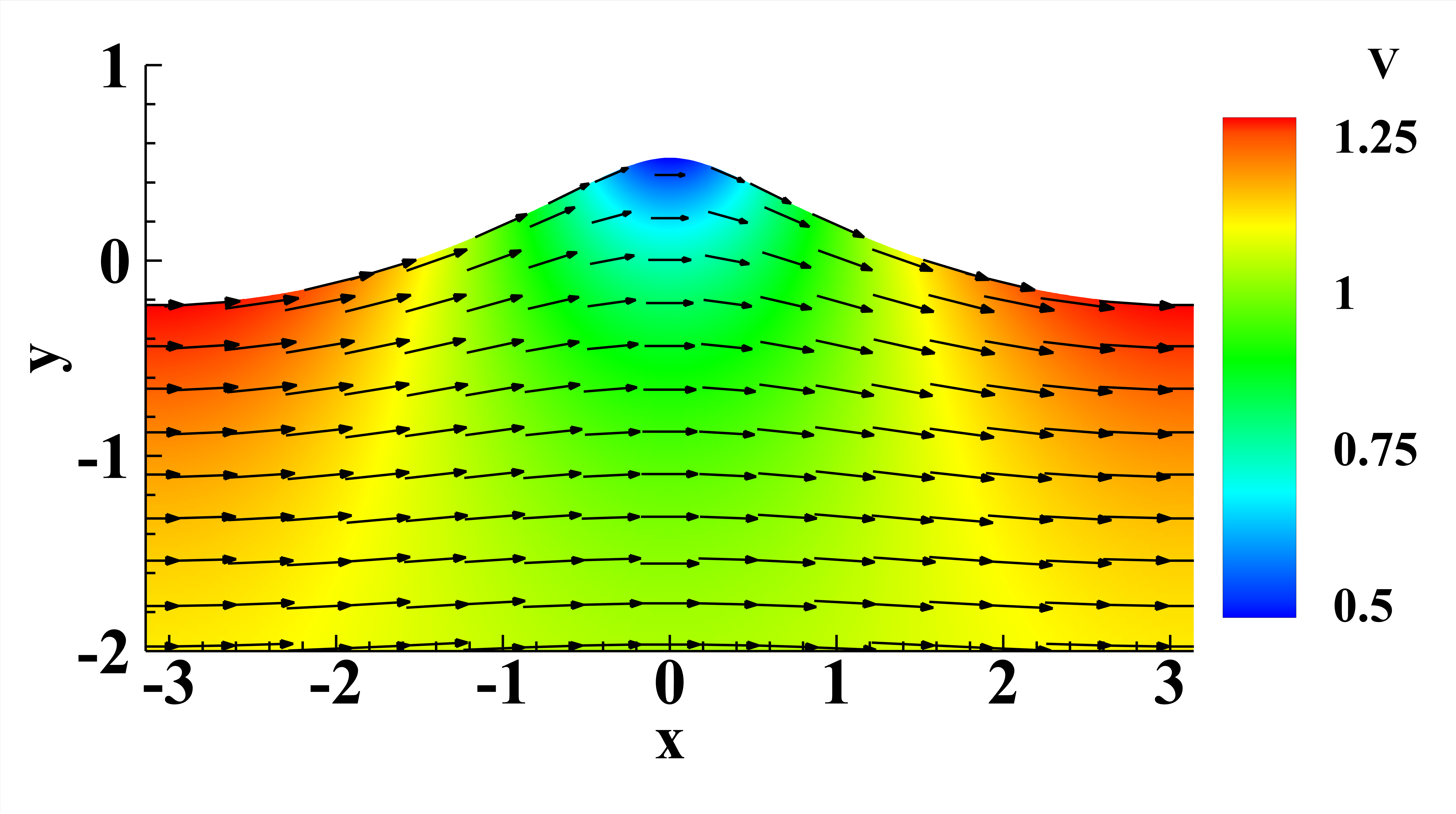}
        \caption{HAM, $A/L=0.120$}
    \end{subfigure}
    \begin{subfigure}{0.3\textwidth}
        \includegraphics[width=\linewidth, trim=10pt 10pt 10pt 10pt, clip]{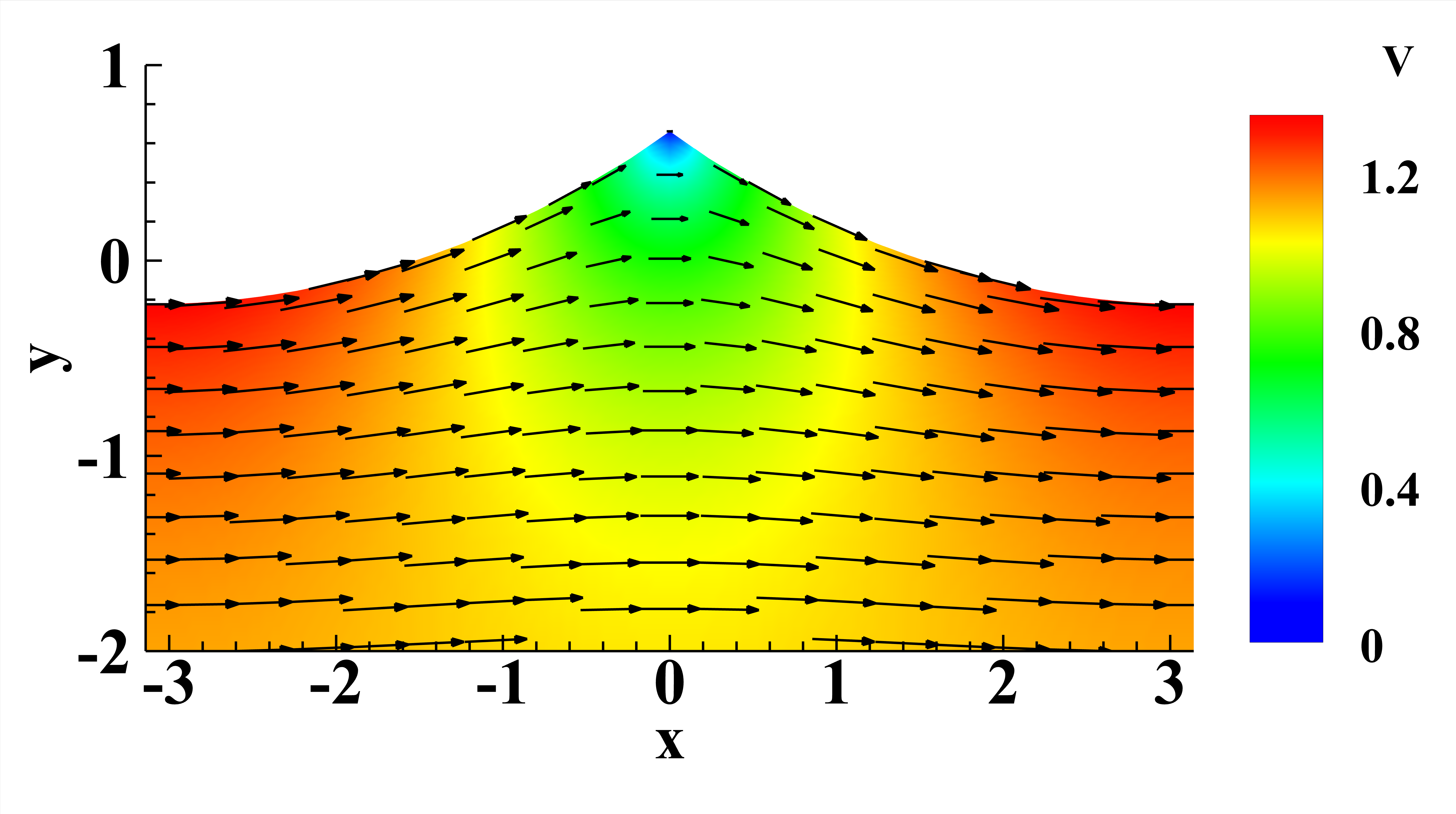}
        \caption{HAM, Limiting wave}
    \end{subfigure}
    \caption{Comparison of wave flow fields using the ML solutions and the benchmark HAM solutions for wave steepness values of $A/L = 0.07$, $0.12$, and the limiting case in infinite-depth water. The top row shows the velocity fields obtained via the machine learning model, while the bottom row presents the corresponding HAM results. Arrows represent the local velocity directions, and color contours indicate velocity magnitudes.}

    \label{fig:ML_aj}
\end{figure}

To evaluate the generalization performance of the trained neural network, we selected $8$ steepness values not included in the train set: $A/L = 0.01$, $0.04$, $0.07$, $0.100$, $0.120$, $0.130$, $0.135$, $0.140$, and $0.141$ as input values. For each input value, the predicted Fourier series generated by the DNN model is then utilized to reconstruct the entire flow field, including the surface profile and velocity distribution,and compared with the benchmark HAM solutions obtained in Section~\ref{sec:HAM-results}. The wave flow fields obtained from the machine learning solutions for wave steepness values of $0.07$, $0.12$ and the limiting case are plotted in Figure~\ref{fig:ML_aj} for comparison with the flow fields obtained from the HAM solutions.

As shown in Figure~\ref{fig:ML_aj}, the two groups of results show excellent consistency in geometry and hydrodynamic characteristics. These results clearly demonstrate that the machine learning model effectively captures both the kinematic and dynamic characteristics of the wave. Notably, even under conditions of strong nonlinearity, characterized by large wave steepness, the trained model exhibits high prediction accuracy.

\begin{figure}[t]  
    \centering
    \includegraphics[width=0.5\textwidth, trim=10pt 10pt 10pt 10pt, clip]{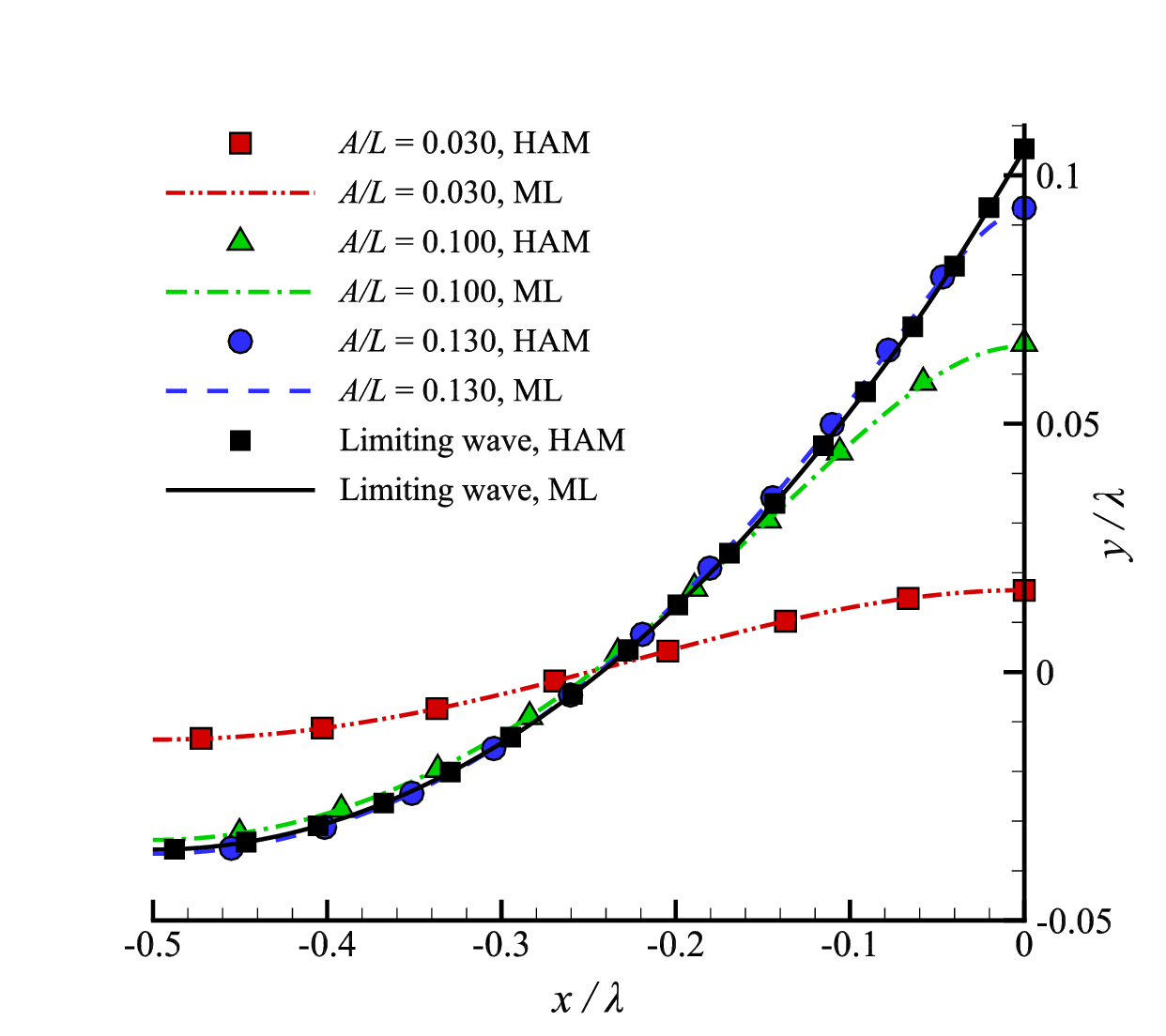}
    \caption{Comparison of free-surface profiles using the ML solutions and the benchmark HAM solutions for different wave steepnesses. Lines, ML solutions; symbols, HAM solutions.}
    \label{fig:comparison_ham_ml}
\end{figure}

To further verify the fidelity of the ML predictions, Figure~\ref{fig:comparison_ham_ml} compares the free-surface profiles obtained via the ML model (lines) with those computed from the HAM (symbols). The two sets of results are visually indistinguishable across a broad range of wave steepness values, including the limiting case. This excellent agreement provides strong evidence for the accuracy and robustness of the ML model.

The statistical error values for the series solutions derived using machine learning are summarized in the Table~\ref{tab:mae_values_aj}. 
Specifically, the coordinates \( (x, y) \) correspond to the physical locations in the fluid domain shown in Figure~\ref{fig:z-plane}, 
which are obtained through the inverse conformal mapping from the \(\zeta\) plane in Figure~\ref{fig:zeta-plane}. 
Both the HAM and ML approaches determine this mapping through the Fourier coefficients \(a_j\) appearing in~\eqref{eq:coordinate-trans}. 
Consequently, small discrepancies between the coefficients predicted by the neural network and those computed via HAM 
lead to slight variations in the mapped physical coordinates \( (x, y) \), thereby introducing coordinate-related errors \(x\) and \(y\)
as listed in Table~\ref{tab:mae_values_aj}. 

In Table~\ref{tab:mae_values_aj}, the subscript HAM denotes the reference solution computed by HAM. Thus, all quantities in Table~\ref{tab:mae_values_aj} are expressed in non-dimensional or relative form. The coordinate deviations are measured on a non-dimensional grid scaled by the characteristic wavelength. The quantities \(v_x/c_{\text{HAM}}\), \(v_y/c_{\text{HAM}}\), and \(v/c_{\text{HAM}}\) denote the relative magnitudes of the horizontal, vertical, and total velocities, respectively, normalized by the corresponding phase velocity given by HAM. The terms $c^2/c_{\text{HAM}}^2$ and $K/K_{\text{HAM}}$ quantify the relative errors in the squared wave velocity \( c^2 \) and the Bernoulli constant \( K \) with respect to the HAM reference values.
All (relative) errors remain on the order of $10^{-3}$ across the tested steepnesses, confirming the high accuracy and reliability of the ML model's predictions.

\begin{table}
    \centering
    \begin{tabular}{cccccccc}
        \toprule
        \textit{A/L} & $x$ & $y$ & $v_x/c_{\text{HAM}}$ & $v_y/c_{\text{HAM}}$ & $v/c_{\text{HAM}}$ & $c^2/c_{\text{HAM}}^2$ & $K/K_{\text{HAM}}$ \\
        \midrule
        0.010 &$ 1.8\times 10^{-3}$&$1.8\times 10^{-3}$&$1.9\times 10^{-3}$&$1.8\times 10^{-3}$&$1.9\times 10^{-3}$&$4.6\times 10^{-4}$&$7.0\times 10^{-4}$\\
        0.040 &$9.1\times 10^{-4}$&$9.1\times 10^{-4}$&$9.1\times 10^{-4}$&$9.0\times 10^{-4}$&$9.0\times 10^{-4}$&$7.2\times 10^{-3}$&$7.4\times 10^{-3}$\\
        0.070 &$2.1\times 10^{-3}$&$2.0\times 10^{-3}$&$2.9\times 10^{-3}$&$2.2\times 10^{-3}$&$3.0\times 10^{-3}$&$6.8\times 10^{-4}$&$4.4\times 10^{-3 }$\\
        0.100 &$3.3\times 10^{-4}$&$3.1\times 10^{-4}$&$8.7\times 10^{-4}$&$4.1\times 10^{-4}$&$9.0\times 10^{-4}$&$6.9\times 10^{-4}$&$4.4\times 10^{-4}$\\
        0.120 &$ 1.2\times 10^{-4}$&$1.2\times 10^{-4}$&$8.7\times 10^{-4}$&$2.2\times 10^{-4}$&$9.0\times 10^{-4}$&$3.5\times 10^{-3}$&$3.8\times 10^{-3}$\\
        0.130 &$ 7.9\times 10^{-5}$&$7.5\times 10^{-5}$&$8.4\times 10^{-4}$&$1.8\times 10^{-4}$&$8.5\times 10^{-4}$&$2.0\times 10^{-3}$&$2.2\times 10^{-3}$\\
        0.135 &$ 3.0\times 10^{-5}$&$3.1\times 10^{-5}$&$1.8\times 10^{-4}$&$8.0\times 10^{-5}$&$1.8\times 10^{-4}$&$2.8\times 10^{-3}$&$2.5\times 10^{-3}$\\
        0.140 &$ 1.7\times 10^{-4}$&$1.6\times 10^{-4}$&$1.8\times 10^{-4}$&$1.8\times 10^{-4}$&$1.7\times 10^{-4}$&$8.5\times 10^{-3}$&$6.2\times 10^{-3}$\\
        0.141 &$ 3.7\times 10^{-5}$&$3.7\times 10^{-5}$&$2.6\times 10^{-4}$&$7.9\times 10^{-5}$&$2.6\times 10^{-4}$&$6.9\times 10^{-4}$&$6.5\times 10^{-4}$\\
        \bottomrule
    \end{tabular}
    \caption{Mean absolute errors of the ML solutions for various physical quantities corresponding to different wave steepness values in infinite-depth water.}
    \label{tab:mae_values_aj}
\end{table}

It is worth reiterating that the cases shown above are not among the 20 parameter sets used for training. The strong agreement observed therefore provides compelling evidence for the model's generalisation performance. These results support the conclusion: by integrating the homotopy analysis method (HAM) with modern machine learning (ML) techniques, we have established a hybrid framework capable of generating highly accurate series solutions for two-dimensional Stokes waves of arbitrary steepness in infinite depth. This approach not only preserves the mathematical rigor of HAM but also addresses its practical limitations in terms of computational cost and complexity. 

A particularly noteworthy outcome of this work is the efficiency of the learning process. Remarkably, only 20 sets of HAM solutions, each corresponding to a distinct wave steepness, are sufficient to train a neural network that can instantly compute series solutions for arbitrary wave heights across the parameter range. Once trained, the model bypasses the need for iterative symbolic computation or numerical solvers, producing analytical expressions in less than one second. In contrast, the computation of Stokes waves in infinite depth for a given wave steepness using HAM on the same processor usually takes several seconds (for very small wave steepness) or even tens of hours (for near-limiting cases). This means that machine learning model can greatly reduce computing time, memory usage and storage overhead, so this model is especially suitable for high throughput or real-time applications.

Crucially, the output of the machine learning model remains an analytical expression rather than a discrete or interpolated numerical approximation. As such, the solution inherits the continuity, differentiability, and exact symbolic structure characteristic of HAM, yet offers a far more compact and interpretable functional form. This makes it highly amenable to post-processing tasks such as symbolic differentiation or embedding in large-scale simulations. Furthermore, error analysis confirms that the learned solution matches the original HAM results with negligible loss of precision, typically within the same order of magnitude. Preliminary tests suggest that accuracy can be further improved by optimizing network architectures or incorporating physics-informed loss functions.

It is worth emphasizing that the steepness of deep water Stokes waves is inherently bounded by \( 0 \leq A/L \leq 0.14108\) \citep{zhong2018limiting}, corresponding to the theoretical limiting wave. As such, the neural network is explicitly trained to interpolate across this full physical range, with representative cases selected near the theoretical limiting steepness to ensure robustness and accuracy even in strongly nonlinear regimes. Extrapolation beyond this interval is not pursued, as it lies outside the physically meaningful solution space.

Thanks to the ability of the Homotopy Analysis Method (HAM) to generate convergent series solutions without extrapolation, including the limiting case, our results capture non-monotonic trends in the phase speed \(c\) and Bernoulli constant \(K\), as well as the $120^\circ$ sharp crest. These features suggest that HAM retains sensitivity to the singular structure inherent in Stokes waves. However, the precise identification of the singularity locations and orders in the complex plane typically requires specialized methods such as conformal mapping or analytic continuation~\citep{longuet-higgins1978theory, crew2016new, lushnikov2016structure}. The neural network in this study is designed for fast and accurate computation of convergent series solutions across arbitrary steepness, rather than for capturing singular features. Enhancing sensitivity to such subtle features would likely require a richer dataset (with finer resolution near the crest) and the introduction of physics-informed loss functions (PINN, proposed by \citet{raissi2019physics}). These improvements are beyond the scope of the current paper but will be explored in future work.

It is important to note, however, that all current results are formulated in the transformed computational domain, denoted as the \(\zeta\) plane. This domain is introduced through a nonlinear conformal mapping, which greatly simplifies the treatment of free-surface boundary conditions. As a result, the velocity and pressure fields obtained thus far are not yet expressed directly in terms of the physical coordinates \((x, y)\). To enable direct physical interpretation and practical application of the results, such as in the analysis of wave-induced loads on structures, an inverse conformal transformation must be applied to map the fields from the \(\zeta\) plane back to the physical \(z\) plane.

\section{Series solution in physical plane for arbitrary physical parameters}\label{sec:inverse_mapping}

Through the preceding sections, we have established an efficient framework for giving series solutions of nonlinear Stokes waves by expressing the wave profile via Fourier series. This approach enables rapid and accurate determination of wave characteristics for arbitrary wave heights. However, a fundamental limitation of the current framework must be acknowledged: the results are defined in the auxiliary conformal plane, denoted as the $\zeta$ plane. In contrast, practical applications often require flow field information in the physical $z$ plane. Specifically, the current method cannot directly evaluate wave quantities from the physical coordinates \((x, y)\). Instead, we must first compute the inverse mapping from \((x, y)\) to the conformal coordinates \((\theta, R)\) in the $\zeta$ plane, and then recover the flow variables. This inverse mapping involves a highly nonlinear transformation, stemming from the initial conformal mapping~\eqref{eq:coordinate-trans} used to simplify the governing equations, and is analytically intractable and computationally expensive.

\begin{figure}[t]
    \centering
    \includegraphics[width=0.9\textwidth]{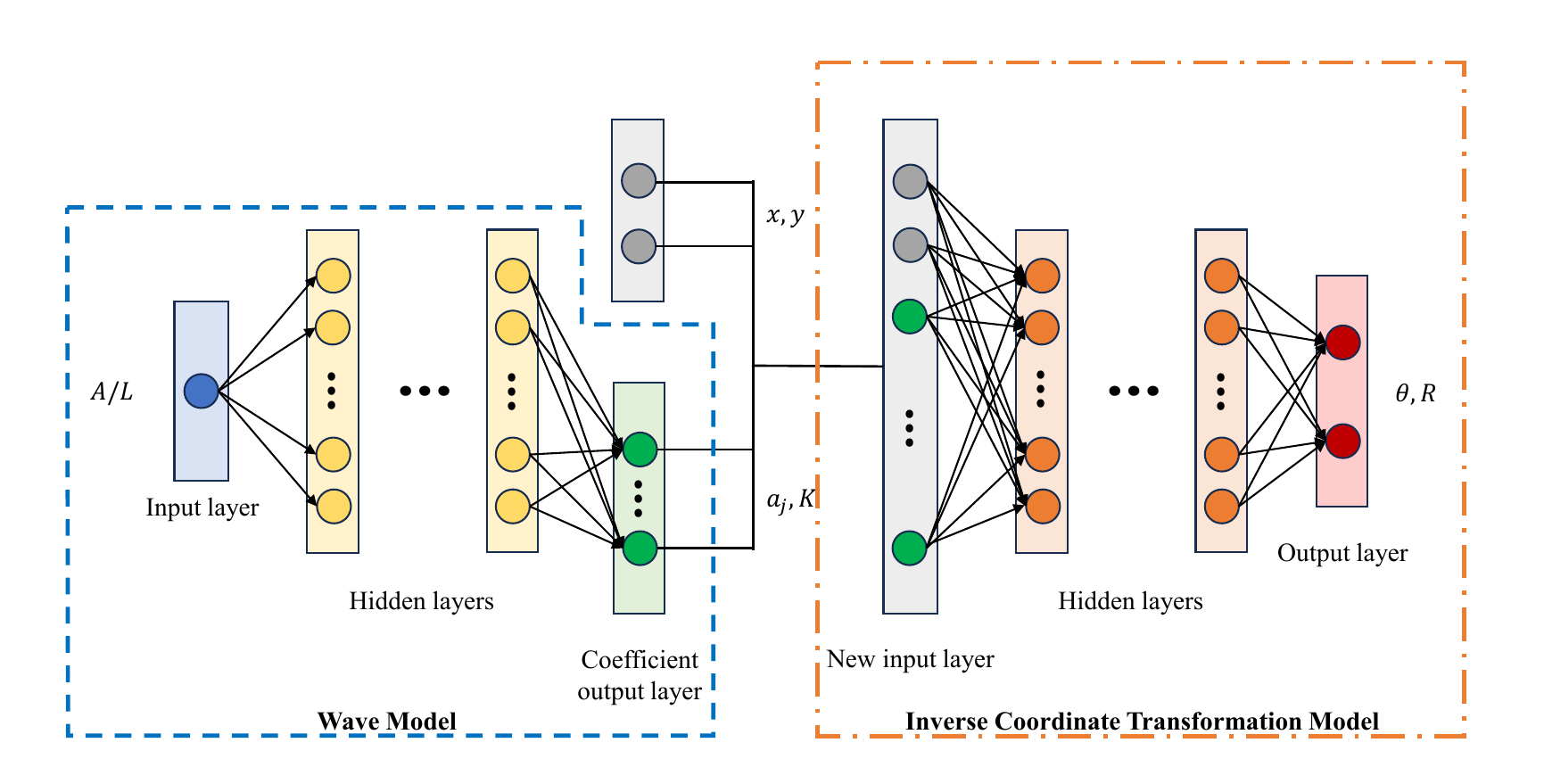}
    \caption{Schematic of the machine learning framework for inverse coordinate transformation. The dashed box indicates the wave model introduced in Section~\ref{sec:ml-results}, while the dash-dotted box denotes the neural network designed for performing the inverse transformation.  The corresponding ML codes can be free downloaded on the website  \url{https://github.com/sjtu-liao/deep-water-Stokes-wave} and \url{https://numericaltank.sjtu.edu.cn/NonlinearWaves.htm} (see ``Convergent Series of Stokes Wave of Arbitrary Height in Deep Water'').}
    \label{fig:DNN-2}
\end{figure}

To overcome the bottleneck of explicitly computing the inverse nonlinear coordinate transformation, we leverage recent advances in machine learning. Neural networks, especially deep architectures, are well-suited to approximate complex, high-dimensional mappings from data. We propose a second neural network model that directly predicts the conformal coordinates \((\theta, R)\) from physical coordinates \((x, y)\). Through this model, the physical \(z\) plane coordinates \((x, y)\) can be efficiently mapped to \(\zeta\) plane coordinates \((\theta, R)\), enabling direct substitution into the series solution formulated under the \(\zeta\) plane. Consequently, this approach effectively yields a series solution expressed in the physical plane without the need for explicit inverse transformations. Recognizing that wave steepness significantly influences the structure of the Fourier series and thus the conformal mapping itself, we design the neural network with strong generalization capabilities. The model inputs both the Fourier coefficients \(a_j\) and the physical coordinates \((x, y)\), and outputs the corresponding conformal coordinates \((\theta, R)\) for the given wave profile. This enables the model to learn a family of nonlinear coordinate inversions across a range of wave steepness values. The schematic of this inverse coordinate transformation model is illustrated in Figure~\ref{fig:DNN-2}, where some inputs are derived directly from the neural network outputs introduced in Section~\ref{sec:ml-results}.  The corresponding ML codes can be free downloaded on the website  \url{https://github.com/sjtu-liao/deep-water-Stokes-wave} and \url{https://numericaltank.sjtu.edu.cn/NonlinearWaves.htm} (see ``Convergent Series of Stokes Wave of Arbitrary Height in Deep Water'').

To further improve the network's ability to learn the feature near the wave crest, we introduce an exponentially weighted sampling strategy. Specifically, we increase sampling density near the wave peak to enhance the model's resolution in strongly nonlinear regions. This data preprocessing technique significantly improves prediction accuracy, especially for waves with high steepness. Using these preprocessed data as input, train the machine learning model using a mean square error (MSE) loss function and an adaptive gradient-based optimizer. Upon training completion, we test the network on waves with steepness values of $0.010$, $0.040$, $0.060$, $0.070$, $0.080$, $0.100$, $0.110$,$0.120$, $0.130$, $0.135$, and $0.141$. It is important to highlight that all the wave cases used for testing and validation are strictly excluded from the training dataset. This design ensures that the evaluation of the deep neural network reflects its true generalization capability, rather than mere interpolation over seen data. For each test case, we calculate and visualize the velocity field, and make point-by-point comparisons between the solution in the physical \(z\) plane after the neural network has completed the inverse coordinate transformation and the solution in the \(\zeta\) plane obtained through HAM. A quantitative error analysis is presented in Table~\ref{tab:mae_values_coo}, where the mean absolute error (MAE) at selected spatial locations is reported. All velocity errors in Table~\ref{tab:mae_values_coo} are reported as relative errors normalized by the phase speed \(c\) computed by HAM. As evidenced by the data in the table, the mean absolute errors of velocity prediction are consistently controlled within the order of \(10^{-4}\) or even lower. This high level of accuracy demonstrates the reliability of the proposed inverse mapping approach in reconstructing the physical flow field from model outputs.

\begin{table}
    \centering
    \begin{tabular}{cccc}
        \toprule
        \textit{A/L} & $v_x/c$ & $v_y/c$ & $v/c$ \\
        \midrule
        0.010&$3.2\times 10^{-5}$&$3.1\times 10^{-5}$&$3.2\times 10^{-5}$\\
        0.040&$1.3\times 10^{-4}$&$1.2\times 10^{-4}$&$1.3\times 10^{-4}$\\
        0.060&$1.6\times 10^{-4}$&$1.2\times 10^{-4}$&$1.5\times 10^{-4}$\\
        0.070&$2.0\times 10^{-4}$&$2.1\times 10^{-4}$&$2.0\times 10^{-4}$\\
        0.080&$3.0\times 10^{-4}$&$2.0\times 10^{-4}$&$3.0\times 10^{-4}$\\
        0.100&$2.6\times 10^{-4}$&$2.6\times 10^{-4}$&$2.6\times 10^{-4}$\\
        0.110&$3.8\times 10^{-4}$&$2.3\times 10^{-4}$&$3.7\times 10^{-4}$\\
        0.120&$3.1\times 10^{-4}$&$3.0\times 10^{-4}$&$3.1\times 10^{-4}$\\
        0.130&$3.3\times 10^{-4}$&$3.2\times 10^{-4}$&$3.3\times 10^{-4}$\\
        0.135&$7.0\times 10^{-4}$&$5.1\times 10^{-4}$&$6.8\times 10^{-4}$\\
        0.141&$1.2\times 10^{-4}$&$1.1\times 10^{-4}$&$1.1\times 10^{-4}$\\
        \bottomrule
    \end{tabular}
    \caption{Mean absolute errors of the predicted velocity fields obtained from the inverse coordinate transformation neural network, evaluated at selected spatial locations across a range of wave steepness values.}
    \label{tab:mae_values_coo}
\end{table}

\begin{figure}[tb!]
    \centering

    \begin{subfigure}{0.3\textwidth}
        \includegraphics[width=\linewidth, trim=10pt 10pt 10pt 10pt, clip]{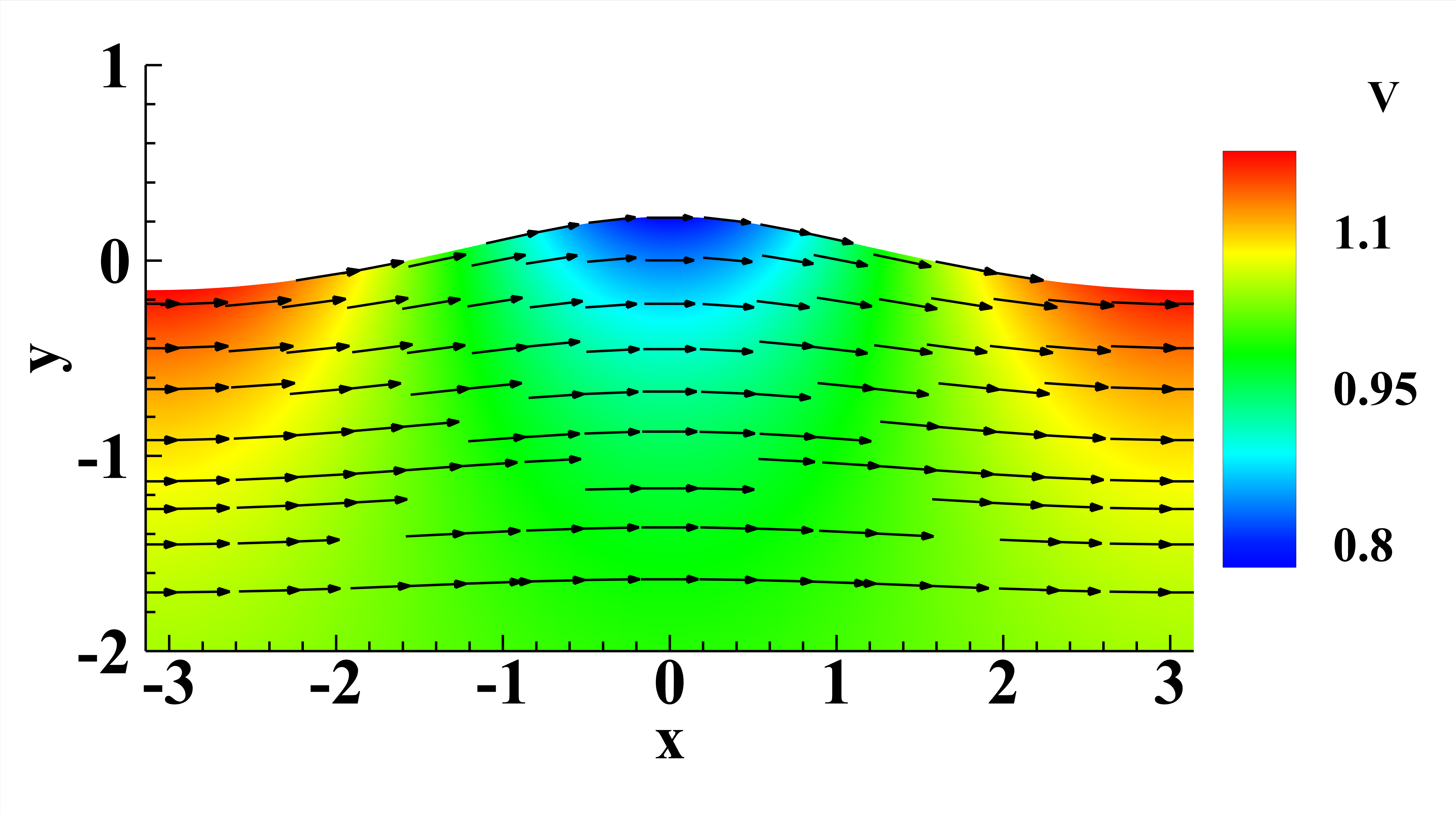}
        \caption{\(\zeta\) plane, $A/L=0.060$}
        \label{fig:040_init}
    \end{subfigure}
    \begin{subfigure}{0.3\textwidth}
        \includegraphics[width=\linewidth, trim=10pt 10pt 10pt 10pt, clip]{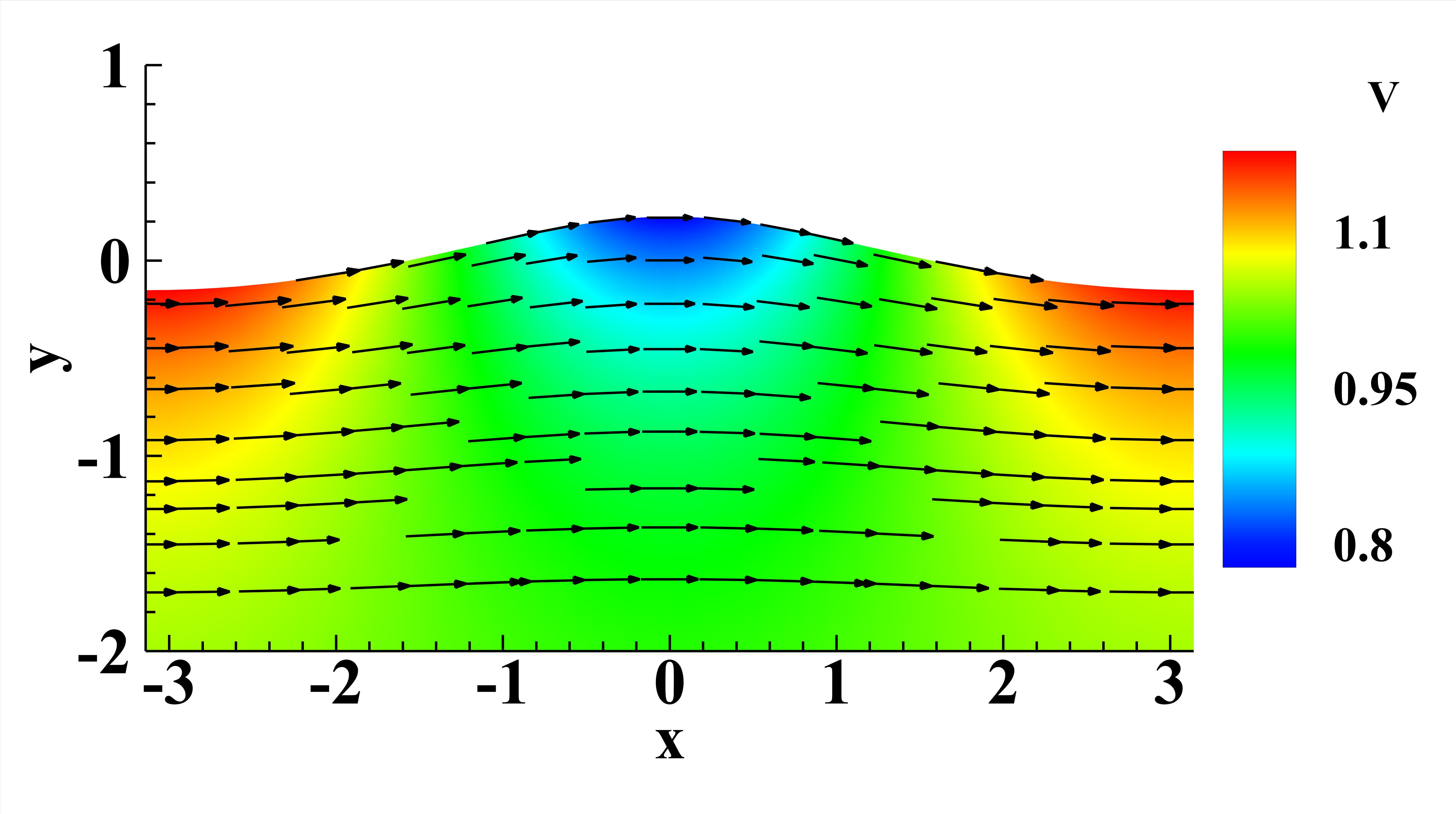}
        \caption{\(z\) plane, $A/L=0.060$}
        \label{fig:040_pre}
    \end{subfigure}
    \begin{subfigure}{0.3\textwidth}
        \includegraphics[width=\linewidth, trim=10pt 10pt 10pt 10pt, clip]{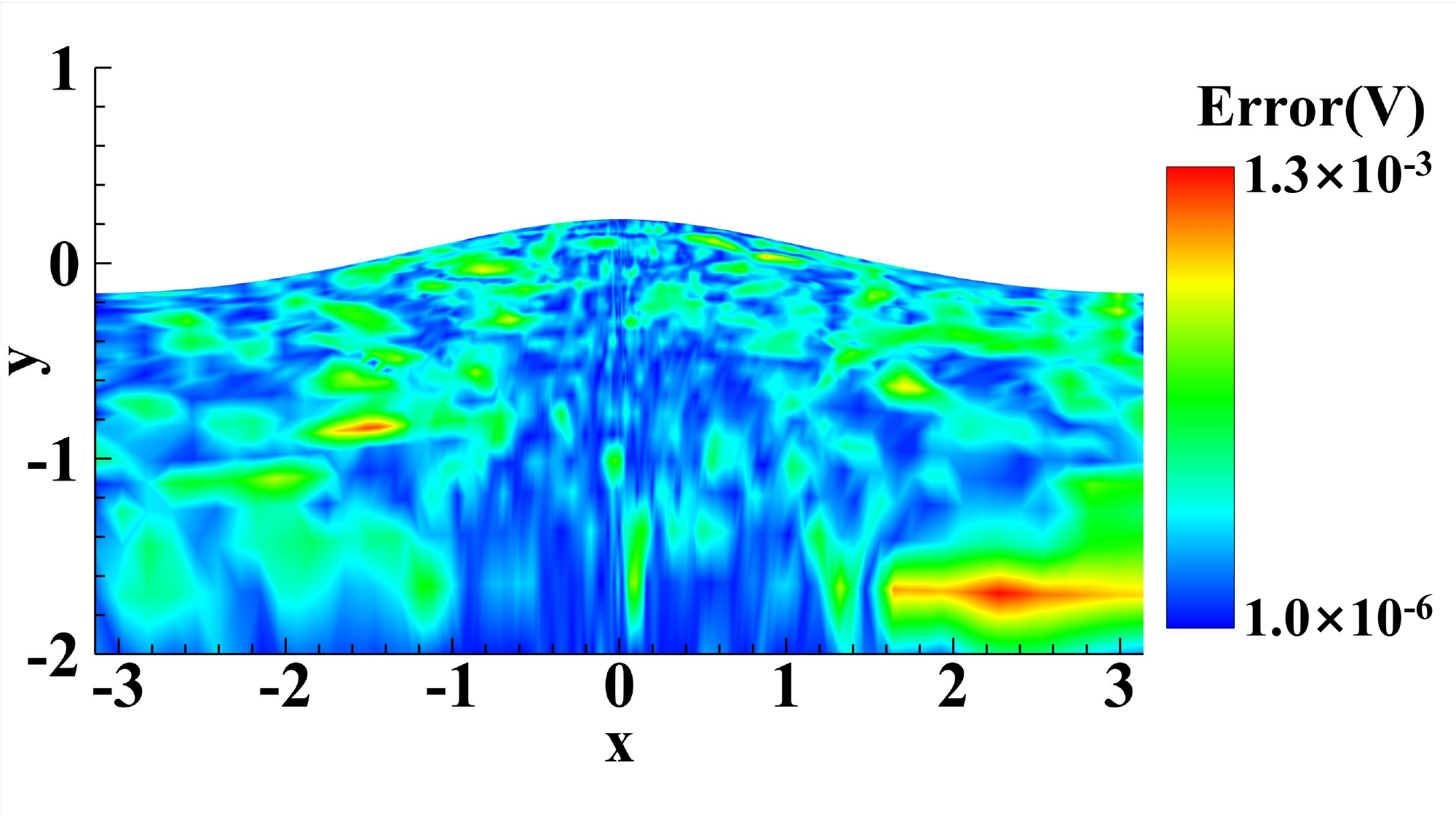}
        \caption{Error, $A/L=0.060$}
        \label{fig:040_error}
    \end{subfigure}

    \begin{subfigure}{0.3\textwidth}
        \includegraphics[width=\linewidth, trim=10pt 10pt 10pt 10pt, clip]{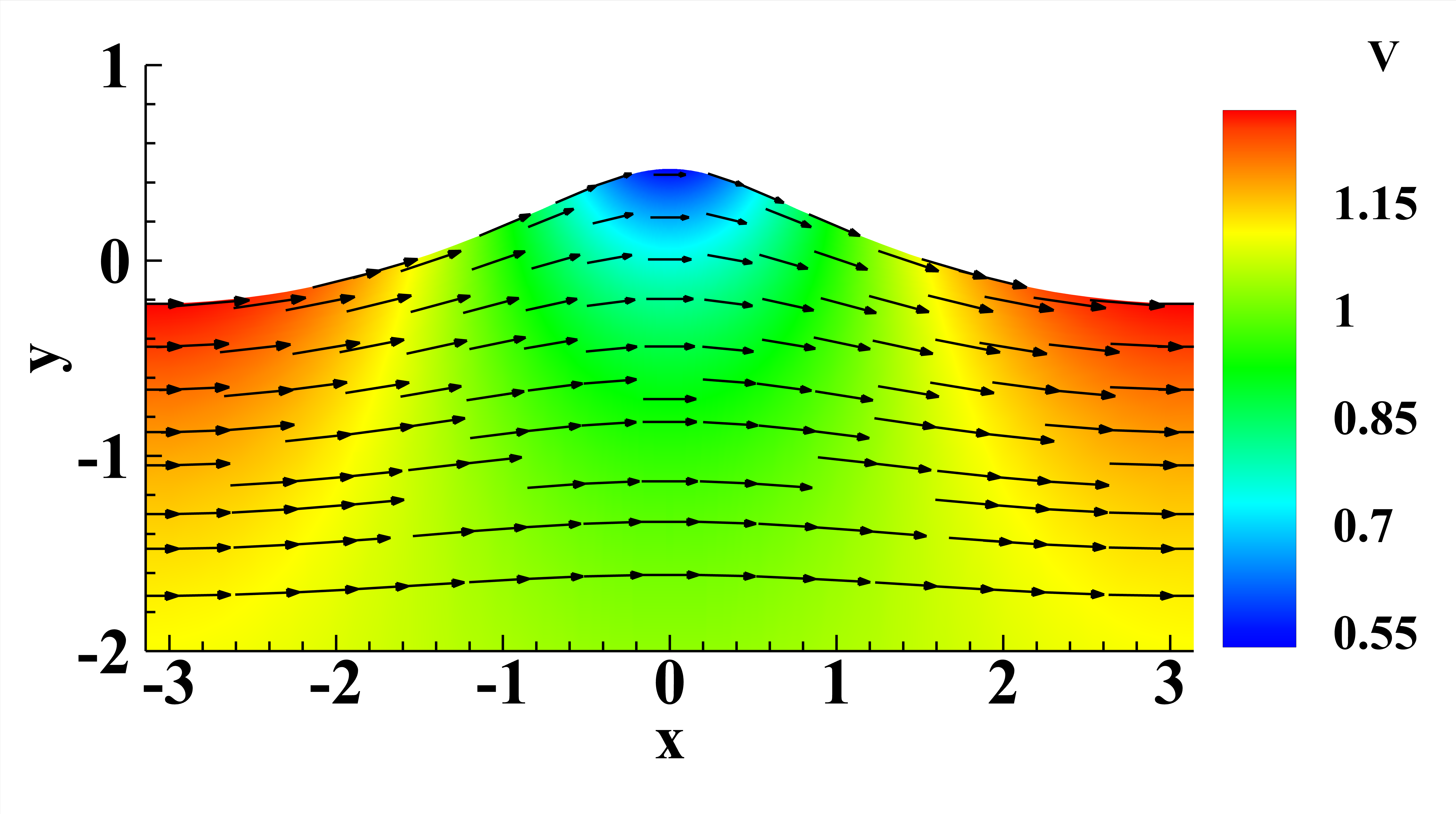}
        \caption{\(\zeta\) plane, $A/L=0.110$}
        \label{fig:100_init}
    \end{subfigure}
    \begin{subfigure}{0.3\textwidth}
        \includegraphics[width=\linewidth, trim=10pt 10pt 10pt 10pt, clip]{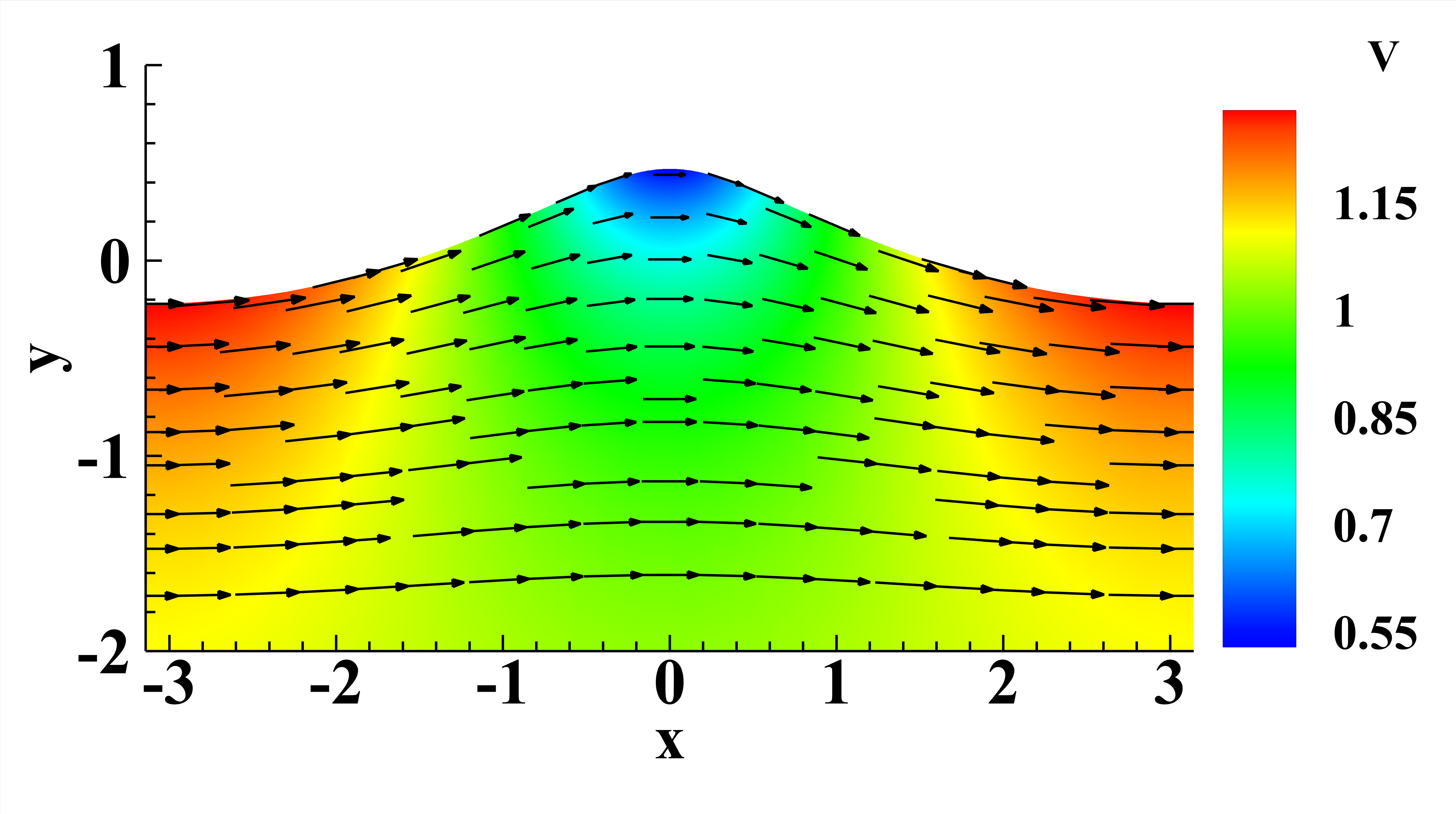}
        \caption{\(z\) plane, $A/L=0.110$}
        \label{fig:100_pre}
    \end{subfigure}
    \begin{subfigure}{0.3\textwidth}
        \includegraphics[width=\linewidth, trim=10pt 10pt 10pt 10pt, clip]{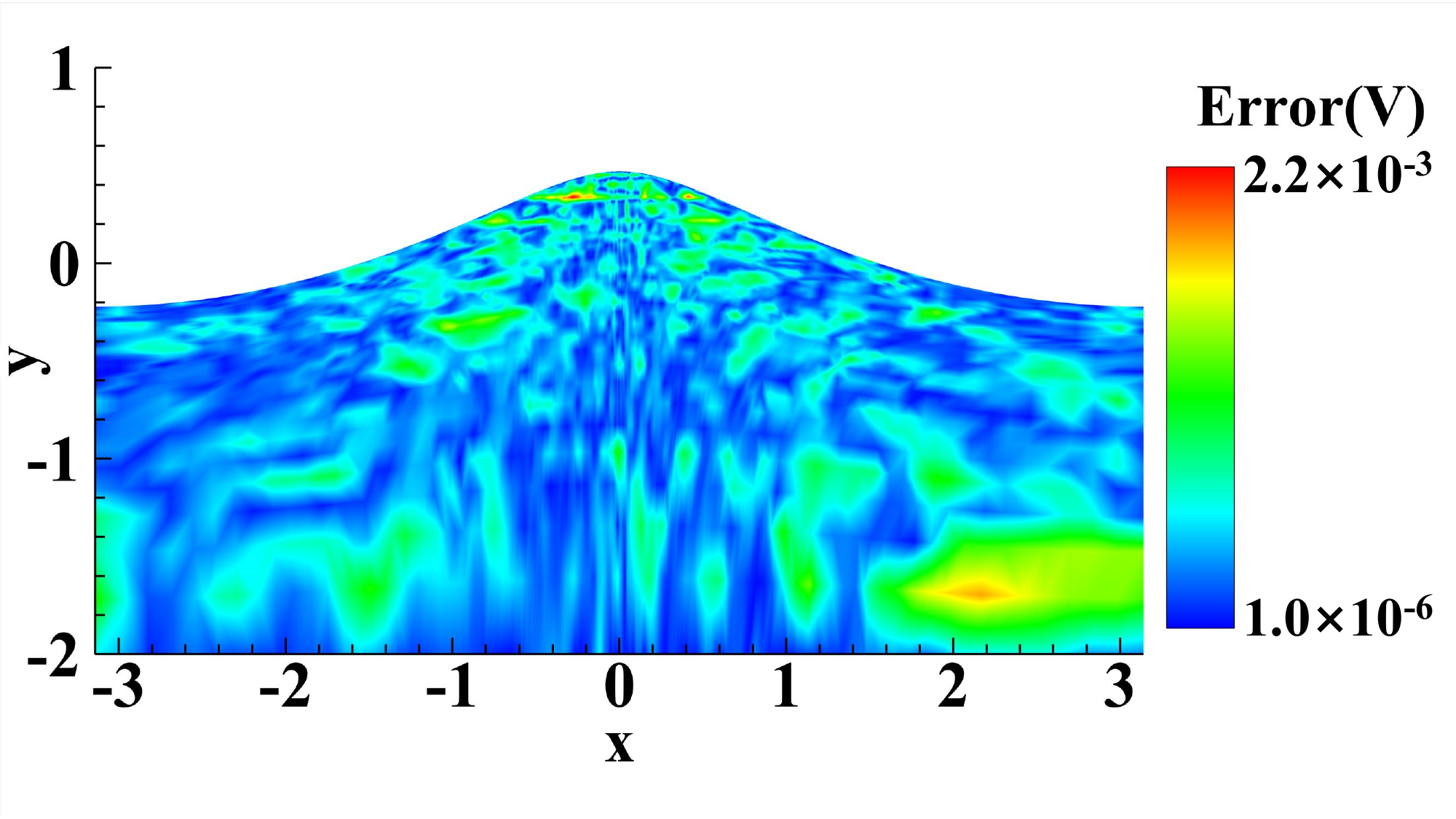}
        \caption{Error, $A/L=0.110$}
        \label{fig:100_error}
    \end{subfigure}

    \begin{subfigure}{0.3\textwidth}
        \includegraphics[width=\linewidth, trim=10pt 10pt 10pt 10pt, clip]{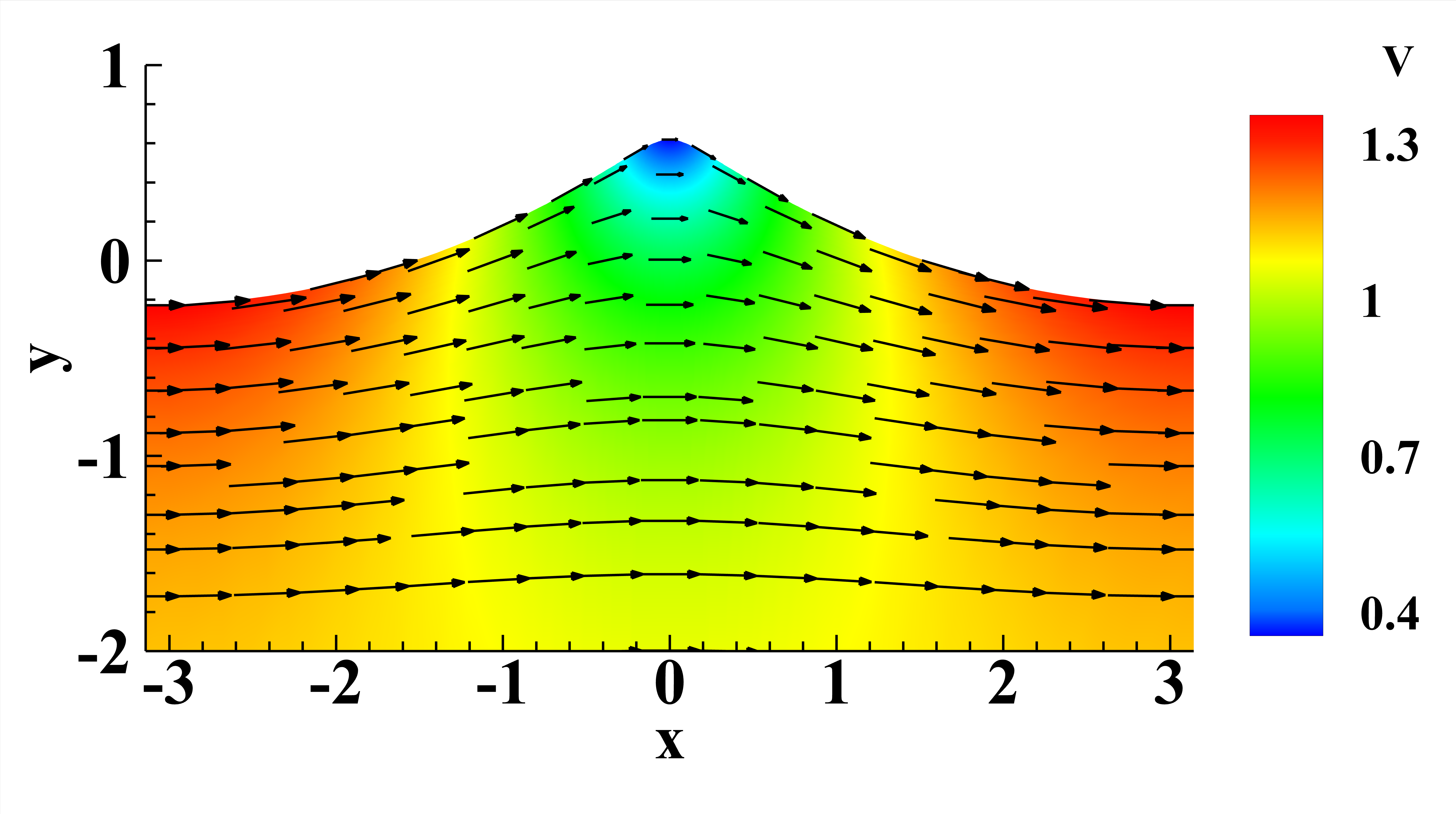}
        \caption{\(\zeta\) plane, $A/L=0.135$}
        \label{fig:130_init}
    \end{subfigure}
    \begin{subfigure}{0.3\textwidth}
        \includegraphics[width=\linewidth, trim=10pt 10pt 10pt 10pt, clip]{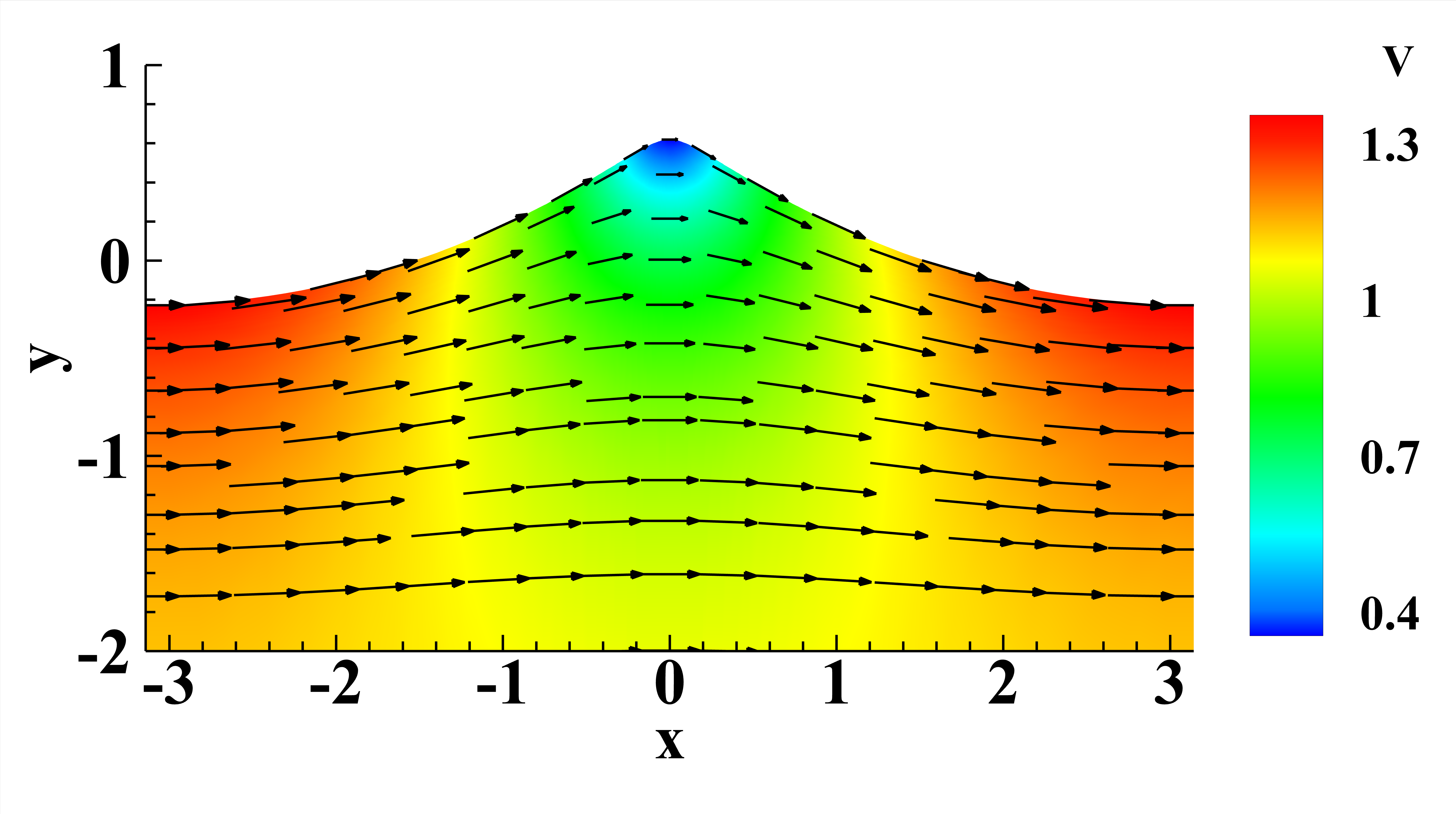}
        \caption{\(z\) plane, $A/L=0.135$}
        \label{fig:130_pre}
    \end{subfigure}
    \begin{subfigure}{0.3\textwidth}
        \includegraphics[width=\linewidth, trim=10pt 10pt 10pt 10pt, clip]{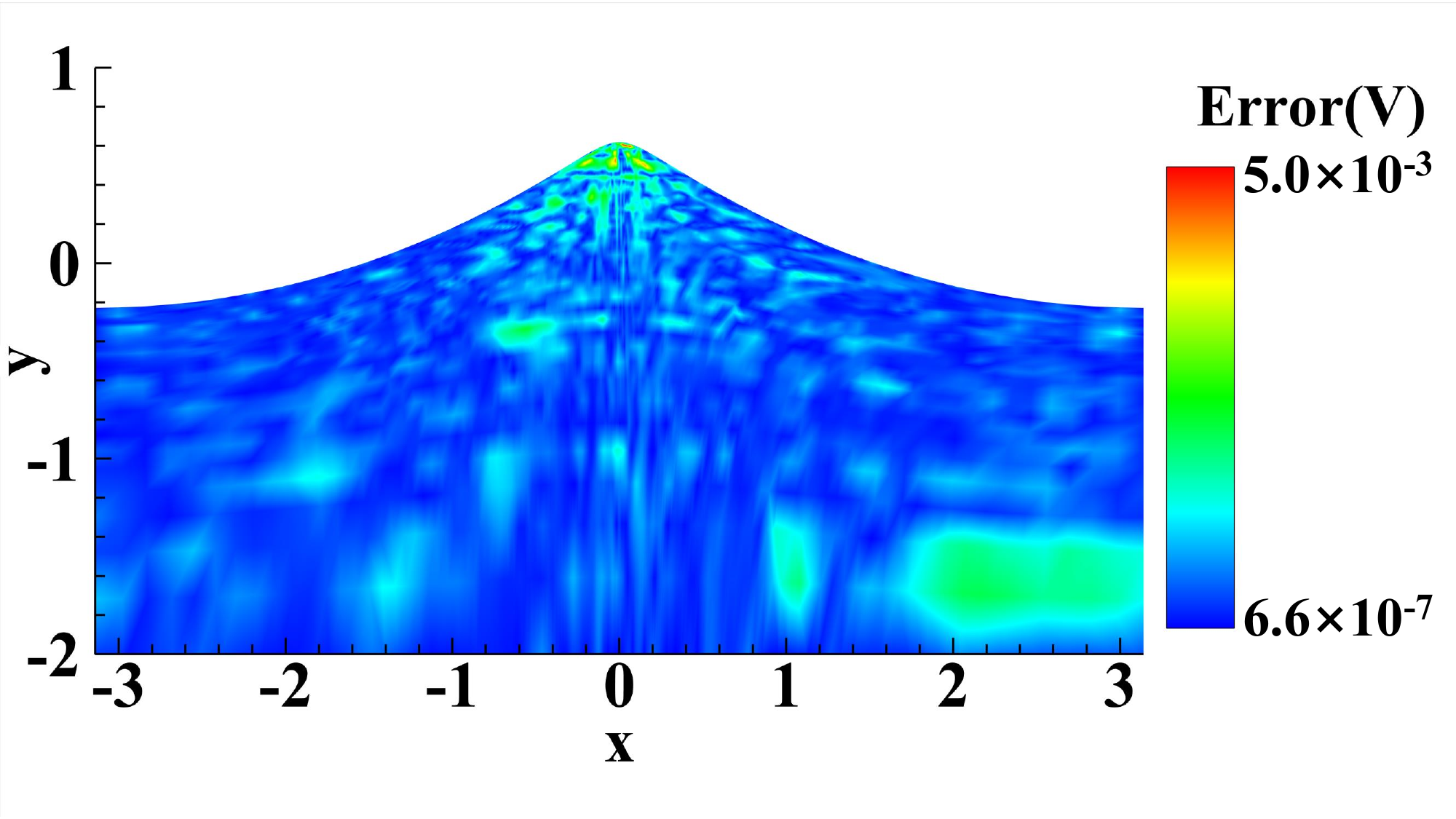}
        \caption{Error, $A/L=0.135$}
        \label{fig:130_error}
    \end{subfigure}
    
    \begin{subfigure}{0.3\textwidth}
        \includegraphics[width=\linewidth, trim=10pt 10pt 10pt 10pt, clip]{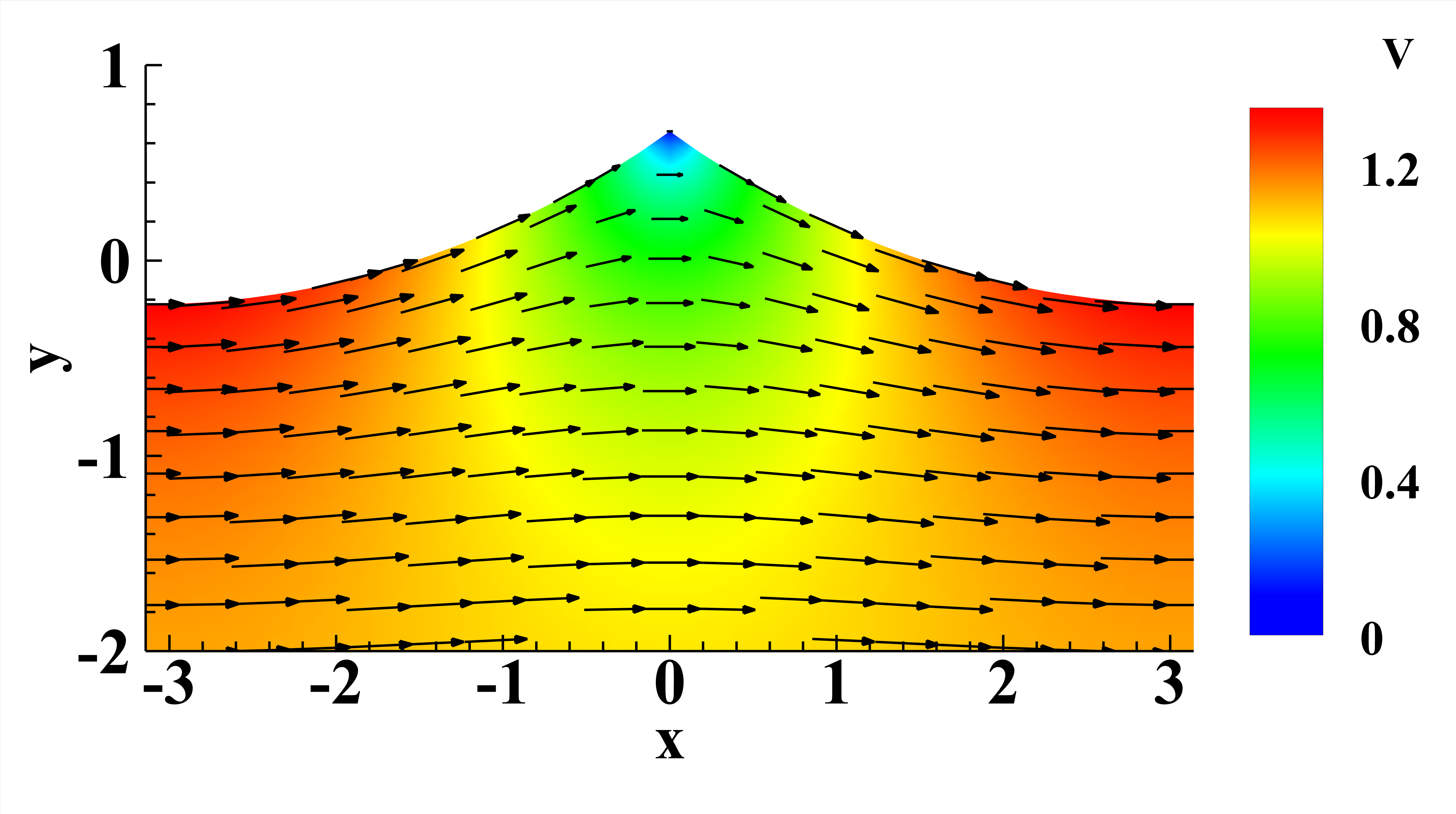}
        \caption{\(\zeta\) plane, Limiting wave}
        \label{fig:141_init}
    \end{subfigure}
    \begin{subfigure}{0.3\textwidth}
        \includegraphics[width=\linewidth, trim=10pt 10pt 10pt 10pt, clip]{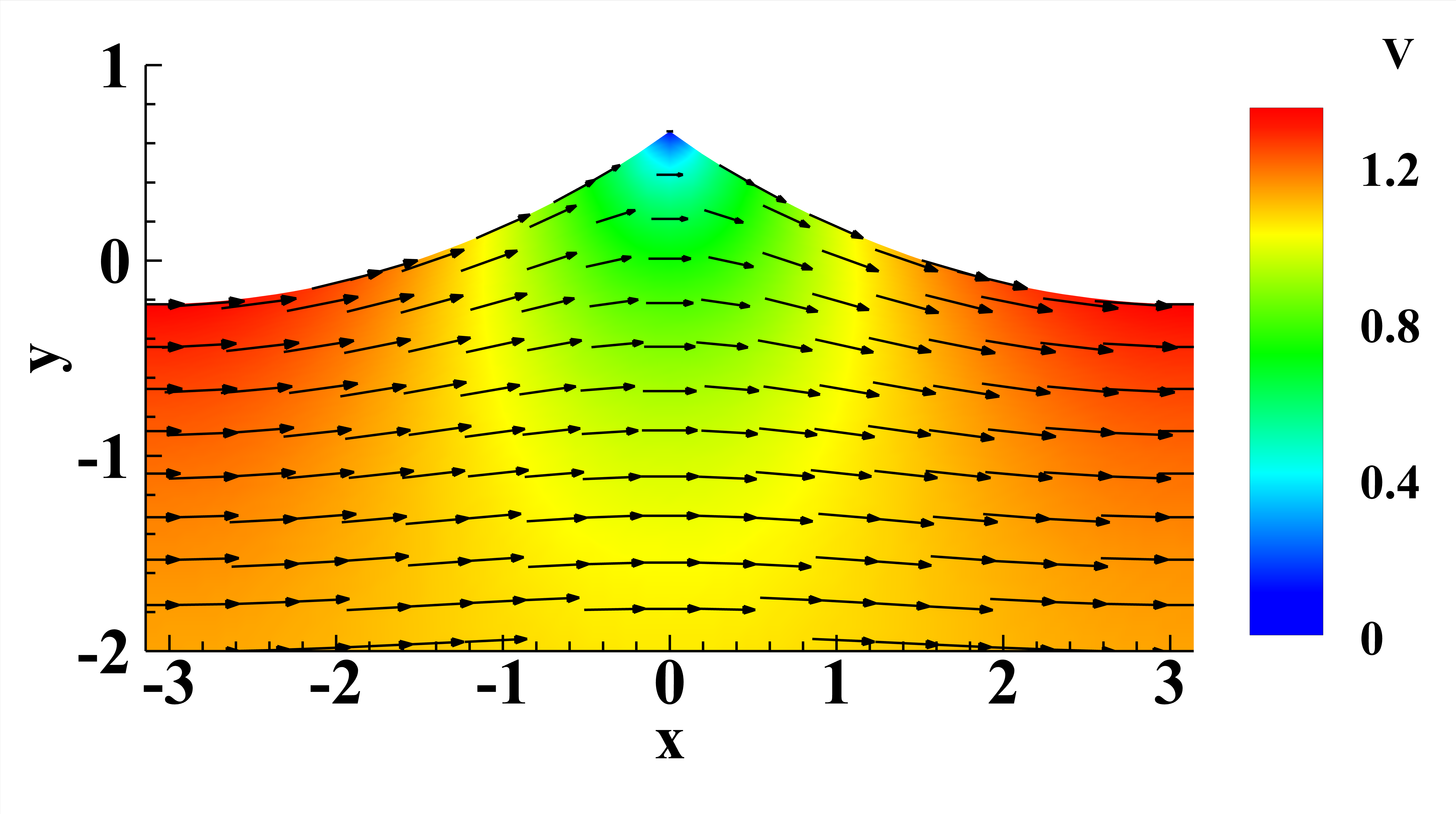}
        \caption{\(z\) plane, Limiting wave}
        \label{fig:141_pre}
    \end{subfigure}
    \begin{subfigure}{0.3\textwidth}
        \includegraphics[width=\linewidth, trim=10pt 10pt 10pt 10pt, clip]{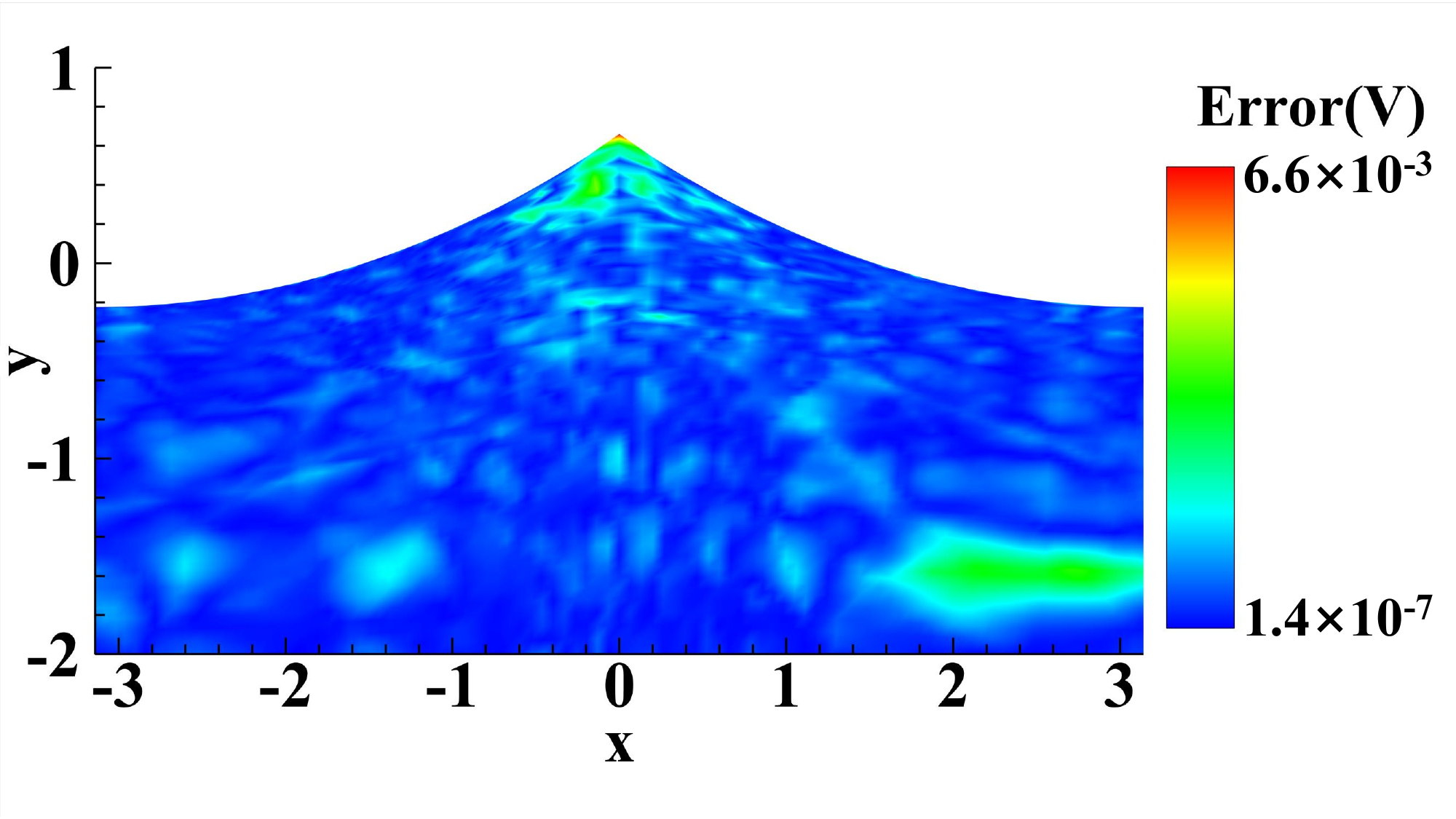}
        \caption{Error, Limiting wave}
        \label{fig:141_error}
    \end{subfigure}
    
    \caption{Comparison of flow fields and corresponding error distributions for selected wave steepness values before and after applying the neural-network-based inverse coordinate transformation.For each case, the left and middle panels display the velocity fields obtained from the HAM solution and those reconstructed via neural network inversion, respectively, while the right panel shows the corresponding pointwise error in velocity magnitude, defined as $\Delta |v| = |v_{\text{NN}} - v_{\text{HAM}}|$. Slightly elevated errors are observed in localized regions with strong nonlinear features or relatively limited spatial resolution, particularly near the wave crest and regions with steep velocity gradients.
}
    \label{fig:ML_coo_with_err}
\end{figure}

We plot the flow field under the solutions in two plane obtained before and after the inverse coordinate transformation, and at the same time plot the error between these two flow fields, as illustrated in Figure~\ref{fig:ML_coo_with_err}. By comparing the predicted flow field reconstructed via neural network inversion with the original flow field derived from the HAM solutions and observing the error distribution, we observe that the overall flow structures are nearly identical. This demonstrates that the proposed inverse mapping approach successfully preserves the key physical features of the solution.

Furthermore, the error contour plots reveal that the prediction error remains extremely low across most regions. In localized areas characterized by strong nonlinearities or relatively low spatial resolution, slightly higher errors are observed. However, these deviations are still well-controlled within the order of \(10^{-3}\), which validates the robustness and accuracy of the method. These findings confirm that the ML-based inverse transformation provides a reliable and efficient means for reconstructing flow fields in the physical plane, even under challenging nonlinear conditions.

\section{Conclusions}\label{sec:conclusion}

Permanent gravity waves propagating in deep water, with amplitudes ranging from infinitesimal to the limiting wave height, represent a classical yet profoundly challenging nonlinear problem.  It is impossible for traditional approaches to {\em quickly} gain series solution convergent for {\em arbitrary} physical parameters having physical meanings.

In this paper  we introduce an innovative and efficient hybrid approach that synergistically combines the convergence guarantee  of the homotopy analysis method (HAM) with the high computational efficiency and flexibility of machine learning (ML). First, we obtain highly accurate, convergent series solutions of deep-water Stokes waves at a {\em finite} set of discrete wave heights by means of HAM, which agree well with the benchmark studies given by~\citet{schwartzComputerExtensionAnalytic1974a}, \citet{dyachenko2016branch} and \citet{zhong2018limiting}. 
Secondly, we extend these series solutions to {\em arbitrary} wave heights using machine learning.  Our neural network, trained on just 20 sets of HAM series solutions, can compute convergent series solutions  within one second for {\em arbitrary} wave height including the limiting cases with very high nonlinearity.  

It is important to note that the traditional series solutions of Stokes waves reside in the conformal \(\zeta\) plane, which are less convenient for practical applications. To overcome this limitation, we introduce another neural network that approximates the inverse conformal mapping  so that any given point \((x,y)\) in the physical plane can be mapped into the \(\zeta\) plane. As a result, all  physical quantities such as the velocity field can be easily reconstructed in the physical plane  \((x,y)\).   
Thus, our framework facilitates the rapid generation of intuitive, convergent series solutions of Stokes waves in deep water, providing initial condition for  various CFD approaches in ocean engineering.

Looking forward, physics-informed neural networks (PINNs) \citep{raissi2019physics} offer a promising alternative or complement to current two networks. By embedding governing equations such as the Laplace and Bernoulli conditions or mapping functions into the training loss function, PINNs could enhance physical consistency, improve resolution near singular features, reduce data requirements, and potentially improve extrapolation behavior. This approach may enable broader extensions to more complex or generalized wave systems.

Traditionally,  it is impossible to obtain a series solution of complicated nonlinear problem like Stokes waves, which are computationally efficient and convergent for {\em all} possible physical parameters.   In this paper, using Stokes wave as an example,  we establish a new {\em paradigm} to  {\em quickly}  obtain convergent series solutions  of complex nonlinear PDEs  valid for {\em all} possible physical parameters,  thereby significantly broadening the scope of convergent series solutions  that can be easily gained  by means of homotopy analysis method (HAM) \citep{liao1992phd, liao2003beyond, liao2009notes, liao2010optimal, liao2011multiple, liao2012ham, Liao2022AAMM, Liao2026AAMM} even for highly nonlinear problems in science and engineering.

\begin{bmhead}[Acknowledgements] 
Thanks a lot to the anonymous reviewers for their helpful comments.   This work is partly supported by National Natural Science Foundation of China (No.  12272230  and  No. 12521002).    
\end{bmhead}

\begin{bmhead}[Declaration of Interests] 
The authors report no conflict of interest.
\end{bmhead}

\begin{bmhead}[Data availability statement] The ML model used in this paper can be free downloaded on the website  \url{https://github.com/sjtu-liao/deep-water-Stokes-wave} and \url{https://numericaltank.sjtu.edu.cn/NonlinearWaves.htm} (see ``Convergent Series of Stokes Wave of Arbitrary Height in Deep Water'').   
\end{bmhead}

\begin{appen}

\section{Detailed derivation of homotopy-series solutions}\label{appA}
Clearly, at $q = 0$, equation~\eqref{eq:HAM_0} has a solution:
\begin{subequations}\label{eq:HAM-q0}
\begin{align}
    K(0) &= K_0, \label{eq:q_0_k} \\
    \Omega_{j}(0) &= a_{j,0} \quad (j=1,2,3, \ldots, r). \label{eq:q_0_a}
\end{align}
\end{subequations}
When \( q = 1 \), the deformation equation~\eqref{eq:HAM_0} becomes:
\begin{subequations}\label{eq:HAM-q1}
\begin{align}
    K(1) &= K, \label{eq:q_1_k}\\
    \Omega_{j}(1) &= a_{j} \quad (j=1, 2, 3, \ldots, r), \label{eq:q_1_a}
\end{align}
\end{subequations}
which coincides with the original nonlinear algebraic system~\eqref{eq:nonlinear_eq}.

Hence, as \( q \) varies continuously from 0 to 1, each function \( \Omega_j(q) \) deforms smoothly from its initial guess \( a_{j,0} \) to the exact unknown coefficient \( a_j \). In the framework of HAM, equation~\eqref{eq:HAM_0} is referred to as the zero-order deformation equation. According to~\eqref{eq:HAM-q0}, we have the Maclaurin series:
\begin{subequations}\label{eq:Maclaurin_series}
\begin{align}
    K(q) &= K_0 + \sum_{k=1}^{\infty} K_{k} q^k,  \label{eq:MAC-a} \\
    \Omega_{j}(q) &= a_{j,0} + \sum_{k=1}^{\infty} a_{j,k} q^k \quad (j=1,2,3, \ldots, r), \label{eq:MAC-b}
\end{align}
\end{subequations}
where,
\begin{subequations}\label{eq:HAM-der}
\begin{align}
    K_{k} &= \mathcal{D}_k[K(q)], \label{eq:Kk} \\
    a_{j,k} &= \mathcal{D}_k[\Omega_{j}(q)] \quad (j=1, 2, 3, \ldots, r),  \label{eq:ajk}
\end{align}
\end{subequations}
in which
\begin{equation}
\mathcal{D}_k[f] = \left. \frac{1}{k!} \frac{\partial^k f}{\partial q^k} \right|_{q=0}
\end{equation}
is called the \( k \)-th order homotopy derivative of \( f \).

It is worth noting that the Maclaurin series~\eqref{eq:Maclaurin_series} depends on the convergence-control parameter \( \hbar \). If \( \hbar \) is chosen such that the series converges at \( q = 1 \), then the homotopy series solutions of the original problem is given by:
\begin{subequations}
\begin{align}
K &= \sum_{k=0}^{\infty}K_{k}, \\
a_j &= \sum_{k=0}^{\infty}a_{j,k} \quad (j=2,3,\ldots,r).
\end{align}
\end{subequations}

Substituting~\eqref{eq:Maclaurin_series} into the zero-order deformation equation~\eqref{eq:HAM_0} and equating like powers of \( q \), we obtain the so-called \( m \)-th order deformation equations:
\begin{subequations}
\begin{align}
K_m - \chi_m K_{m-1} &= \hbar \mathcal{D}_{m-1} [ \mathcal{N}_1 ], \\
a_{j-1,m} - \chi_m a_{j-1,m-1} &= \hbar \mathcal{D}_{m-1} [ \mathcal{N}_{j} ] \quad (j=2,3,\ldots,r),
\end{align}
\end{subequations}
where
\begin{align}
\mathcal{D}_i [\mathcal{N}_j] = \sum_{n=0}^{i} \left[ \sum_{l=1}^{r} \frac{a_{l,i-n}}{l} \left\{f_{\lvert l-j \rvert, n} + f_{j+l, n}\right\} - K_{i-n}f_{j,n} \right] \quad (j=1,2,\ldots,r),\label{eq:HAM_DK}
\end{align}
in which
\begin{equation}
\chi_m = 
\begin{cases} 
0 & \text{when } m \leqslant 1, \\ 
1 & \text{when } m > 1 ,
\end{cases}
\end{equation}
Among them, $f_{j,i}$ is a function of $a_{j,i}$, which can be regarded as an intermediate variable introduced to simplify the computation. Its explicit expression is given by:
\begin{subequations}
\begin{align}
f_{0,i} &= (1-\chi_{i+1}) + \sum_{n=0}^{i} \sum_{l=1}^{r} a_{l,i-n} a_{l,n}, \label{eq:f0i} \\
f_{1,i} &= a_{1,i} + \sum_{n=0}^{i} \sum_{l=1}^{r} a_{l,i-n} a_{l+1,n}, \label{eq:f1i} \\
f_{j,i} &= a_{j,i} + \sum_{n=0}^{i} \sum_{l=1}^{r} a_{l,i-n} a_{l+j,n}. \quad (j=2,3,\ldots,r) \label{eq:fji}
\end{align}
\end{subequations}

One significant advantage of the HAM framework is the freedom to choose the initial guesses $K_0$, $a_{1,0}$, $a_{2,0}$, $\ldots$, $a_{r,0}$. In this work, to ensure both physical relevance and rapid convergence, we select the converged solution of \citet{zhong2018limiting} in the deep-water limiting case (\( r_0 = 0 \)), which is known to satisfy the governing equations accurately. Once the initial guesses are selected, the higher-order terms \( K_k, a_{1,k}, a_{2,k}, \ldots, a_{r,k} \) can be computed recursively from~\eqref{eq:HAM_DK}, starting from \( k = 1 \). The \( n \)-th order homotopy approximation of the solution is then written as:
\begin{subequations}
\begin{align}
\tilde{K}_n &= \sum_{k=0}^{n}K_{k}, \\
\tilde{\Omega}_{j,n} &= \sum_{k=0}^{n}a_{j,k} \quad (j=1,2,3,\ldots,r) .
\end{align}
\end{subequations}

\section{Comparison with the fifth-order Stokes expansion}\label{appB}
To further evaluate the accuracy and applicability of the proposed machine learning (ML) model, we compare ML solutions with those of the classical fifth-order Stokes expansion~\citep{fenton1985fifth}, a widely used analytical approximation for periodic surface gravity waves in deep water.

Following the formulation of \citet{fenton1985fifth}, the free-surface elevation \(\eta(x)\) in deep water can be expressed as a power series in terms of the dimensionless wave amplitude \(\epsilon = kA/2\), where \(k\) is the wavenumber and \(A\) is the wave height:
\begin{align}
  k \eta(x) = & \ \epsilon \cos kx + \frac{1}{2} \epsilon^2 \cos 2kx + \frac{3}{8} \epsilon^3 (\cos 3kx - \cos kx) \nonumber \\
  & + \frac{1}{3} \epsilon^4 (\cos 2kx + \cos 4kx) + \frac{1}{384} \epsilon^5 (-422 \cos kx + 297 \cos 3kx \nonumber \\
  & + 125 \cos 5kx) + O(\epsilon^6),
  \label{eq:stokes5}
\end{align}

We perform a direct comparison between wave profiles computed by the ML model and those computed using Equation~\eqref{eq:stokes5} across a range of wave steepnesses. Representative results are shown in Figure~\ref{fig:stokes5}, where hollow symbols indicate ML solutions and filled symbols indicate fifth-order Stokes solutions. 

\begin{figure}[h!]
  \centering
  \includegraphics[width=0.5\textwidth, trim=10pt 10pt 10pt 10pt, clip]{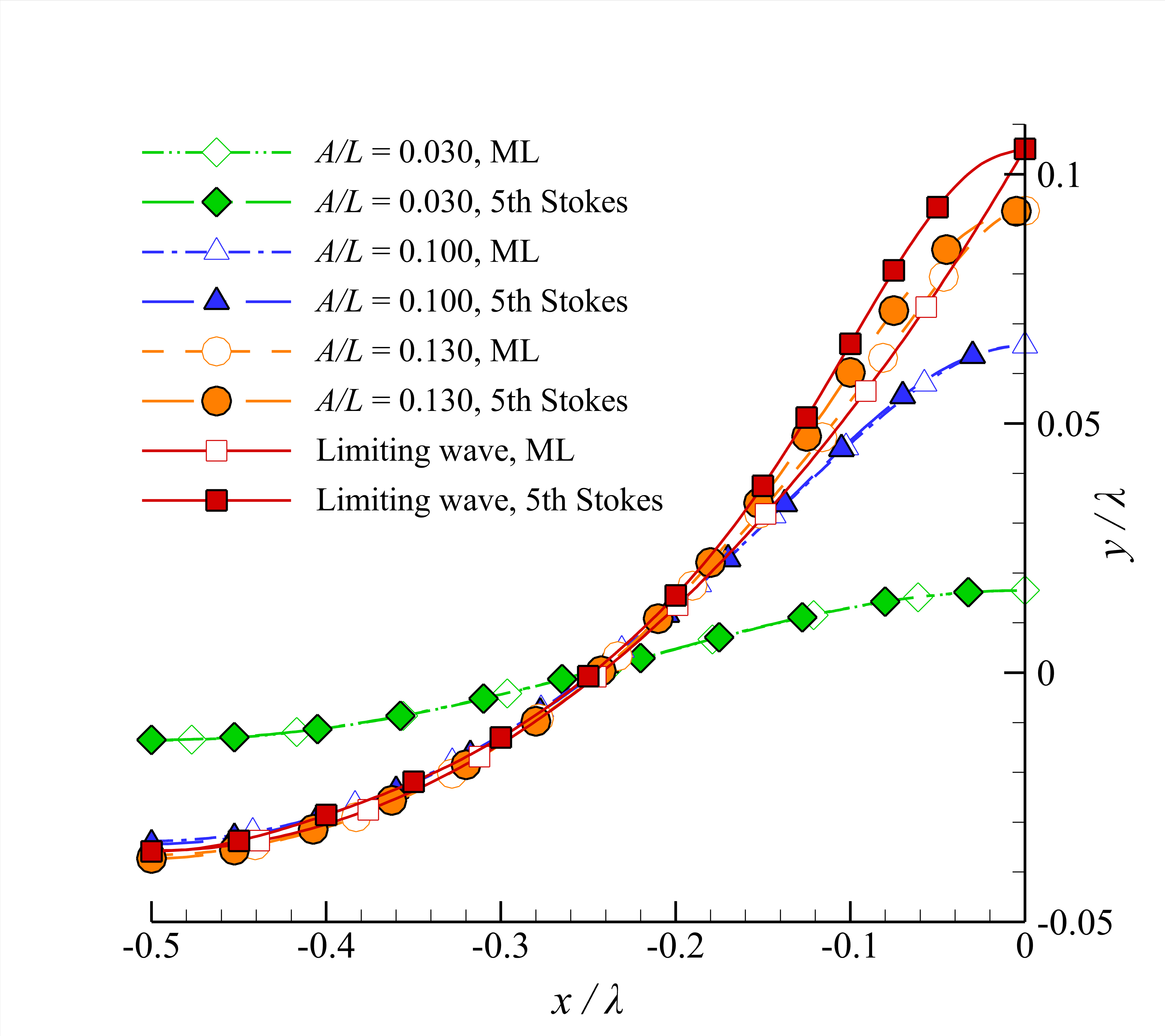}
  \caption{Comparison of free-surface profiles using the ML solutions and the 5th-order Stokes expansion solutions for different wave steepness values.}
  \label{fig:stokes5}
\end{figure}

The results indicate that for weakly nonlinear waves (low steepness), the ML solution agrees closely with the 5th-order Stokes expansion. However, as the steepness increases, the 5th-order Stokes expansion begins to deviate noticeably from the ML solution, particularly near the wave crest, where the profile given by 5th-order Stokes expansion becomes overly rounded rather than sharp. In the limiting case, the 5th-order Stokes expansion fails to capture the characteristic $120^\circ$ crest angle of the highest wave, and the resulting profile is no longer physically accurate. 

This comparison supports the accuracy and generalizability of the ML model from weakly to strongly nonlinear conditions and demonstrates superiority over the fifth-order Stokes approximation outside the weakly nonlinear regime.

\end{appen} 

\bibliographystyle{jfm}
\bibliography{jfm}

\end{document}